\documentclass[namedreferences,hyperref,optionalrh]{spr-sola}
\usepackage{graphicx}        
\usepackage{color}           
\usepackage{epstopdf}
\usepackage{tikz}
\usepackage{rotating}

\chardef\us=`\_

\begin{document}

\begin{frontmatter}
\title{Ion acceleration in Fermi-LAT behind-the-limb solar flares: the role of coronal shock waves}

\author[addressref={aff1},email={awarmuth@aip.de}]{\inits{A.}\fnm{Alexander}~\snm{Warmuth}\orcid{0000-0003-1439-3610}}
\author[addressref=aff2]{\inits{M.}\fnm{Melissa}~\snm{Pesce-Rollins}\orcid{0000-0003-1790-8018}}
\author[addressref=aff3]{\inits{N.}\fnm{Nicola}~\snm{Omodei}\orcid{0000-0002-5448-7577}}
\author[addressref={aff1,aff4}]{\inits{S.}\fnm{Song}~\snm{Tan}\orcid{0000-0003-0317-0534}}
\address[id=aff1]{Leibniz-Institut f\"ur Astrophysik Potsdam (AIP), An der Sternwarte 16, D-14482 Potsdam, Germany}
\address[id=aff2]{Istituto Nazionale di Fisica Nucleare, Sezione di Pisa, I-56127 Pisa, Italy}
\address[id=aff3]{W.W. Hansen Experimental Physics Laboratory, Kavli Institute for Particle Astrophysics and Cosmology, Department of Physics and SLAC National Accelerator
Laboratory, Stanford University, Stanford, CA 94305, USA}
\address[id=aff4]{Institut für Physik und Astronomie, Universität Potsdam, Karl-Liebknecht-Straße 24/25, 14476 Potsdam, Germany}
\runningauthor{A. Warmuth et al.}
\runningtitle{Coronal shocks and ion acceleration in Fermi-LAT behind-the-limb flares}

\begin{abstract}

We investigate the relationship between the gamma-ray emission measured with Fermi-LAT and radio signatures of coronal shock waves in four behind-the-limb (BTL) solar flares. All events were associated with metric type~II radio burst. Both start and end times of the radio bursts were synchronized with the gamma-ray emission. The type II bursts associated with the BTL gamma-ray flares had higher
speeds and lower formation heights than those of an average sample. These findings support the notion that the highly relativistic ions that produce the gamma-rays in BTL flares are accelerated at CME-driven propagating coronal shock waves rather than in large-scale coronal loops.

\end{abstract}
\end{frontmatter}

\section{Introduction}
\label{sec:intro} 

In solar eruptive events - flares and coronal mass ejections (CMEs) - charged particles are accelerated to nonthermal energies. Solar energetic particles (SEPs) escaping into interplanetary space can be directly measured in-situ by spacecraft, while particles precipitating onto the Sun can be constrained by remote-sensing observations, namely by hard X-rays (HXR) for electrons \citep[e.g.][]{Holman2011}, and gamma-rays for ions \citep[e.g.][]{Vilmer2011}. In addition, energetic electrons also emit at radio wavelengths, predominantly via the gyrosynchrotron and plasma emission mechanisms \cite[e.g.][]{White2011}.

It is well-established that energetic electrons and ions contain a significant fraction of the energy released in solar flares \citep[e.g.][]{Emslie2012,Warmuth2020}. Particle acceleration precedes the heating of the bulk of the flare plasma in time and is thus considered as the more direct consequence of the energy release that is presumably triggered by magnetic reconnection. Although the observational constraints on electrons in solar flares are much tighter than on ions, it is generally assumed that the total energy content of ions with energies above $\approx$1~MeV is comparable to the energy in nonthermal electrons \citep[e.g.][]{Shi2009}, and that their acceleration is generally synchronous in time \citep[e.g.][]{Lin2003}.

This scenario has been challenged by the observation of high-energy ($>$100~MeV) gamma-ray emission that can have a long duration~\citep{ryan00,Share_2018,flarecatalog_2021}, or is even associated with behind-the-limb (BTL) flares~\citep{2017FermiBTL}, as observed by Fermi-Large Area Telescope~\citep[LAT;][]{LATPaper}. The latter scenario requires that high-energy ions accelerated in the solar eruptive event have access to the lower atmospheric layers on the Earth-facing hemisphere. Two scenarios have been proposed to account for this. Either flare-accelerated particles could be injected into large-scale magnetic loops connecting to the visible disk~\citep[see for example][]{ryanlee91,1992ApJ...396L.111M,Lit:Som-95,ryan00,Her:al-02}. Alternatively, the ions could be accelerated at an extended CME-driven shock wave~\citep[see for example][]{rank2001,cliv93,Vestrand1993,Plotnikov_2017,Gopalswamy_2019,Jin_2018,Kouloumvakos_2020}.

Recently, we have reported evidence for the shock wave scenario. We have shown that in a BTL flare on 2021 Jul 17~\citep{Pesce-Rollins_2022} the gamma-ray emission was delayed by $\approx$15~min from the impulsive hard X-ray (HXR) emission recorded by the the Spectrometer/Telescope for Imaging X-rays~\citep[STIX;][]{2020A&A...642A..15K} instrument on Solar Orbiter, which does not support a flare-related origin of the energetic ions. However, we found that the gamma-ray emission is synchronized with the appearance of a coronal wave at the visible disk. Such waves are interpreted as signatures of fast-mode MHD waves or shocks.

We also showed that the gamma-ray emission was consistent with the appearance of coronal waves in three other BTL flares \citep{Pesce-Rollins_2022}. However, large-scale coronal waves are not necessarily signatures of shock waves. Based on kinematical characteristics, it has been shown that there exist at least two categories of coronal waves: small-amplitude (i.e., linear) waves that propagate at the ambient fast-mode speed, and large-amplitude (nonlinear) disturbances that travel in excess of the local fast-mode speed and may be (at least partially) shocked \citep{2011A&A...532A.151W,muhr2014}. Estimates for the fast-mode speed in the quiet corona are on the order of a few 100~km\,s$^{-1}$, and two of the four coronal waves discussed in \cite{Pesce-Rollins_2022} had speeds in this range.

In contrast, type~II solar radio bursts that are observed as drifting emission bands in dynamic radio spectra \citep[][]{wild1950} provide unambiguous evidence for the presence of shock waves in the corona since they are generated by shock-accelerated electrons that excite Langmuir turbulence, which is in turn converted to electromagnetic emission \citep[e.g.][]{mann2018}. Indeed, in the BTL flare on 2021 Jul 17 a complex type~II burst was observed in temporal coincidence with the gamma-ray emission, significantly strengthening the case for a shock association of the accelerated ions.

 While a comprehensive study has not yet been performed, we have found similar results also for on-disk gamma-ray flares. For example, the events of 2011 Sep 9 and 2012 June 3 both have onsets in coincidence with the first appearance of the associated coronal waves. We leave the analysis of the on-disk flares for a follow-up paper, but we note  that \cite{Gopalswamy2018} have found a good connection between the high-energy gamma-ray emission in on-disk events and interplanetary type~II radio bursts.

There has been a single case observed so far that shows evidence of flare-accelerated particles in large-scale loops for a BTL gamma-ray flare, thus  providing support for the alternative scenario for $>$100~MeV emission \citep{Pesce-Rollins_2024}. No coronal wave was observed in association with this event and the gamma-ray time profile was impulsive, thus distinctly different from all the other high-energy BTL flares observed to date.

In this paper, we examine dynamic radio spectra for the other BTL events where coronal waves were observed for evidence of coronal shock waves. For completeness, we include the event already analyzed in detail by \cite{Pesce-Rollins_2022} here as well. In Sect.~\ref{sec:obs} we describe the observations and data analysis techniques used. The results are presented in Sect.~\ref{sec:res} , and the conclusion is given in Sect.~\ref{sec:conc}.

\section{Observations and Data Analysis}
\label{sec:obs}

We study four BTL flares for which Fermi-LAT detected gamma-ray emission. The flare dates are listed in Table~\ref{tab:flareinfo}, together with the flare locations (in Stonyhurst coordinates) and occultation angles (defined as the minimum angular distance between the flare and the solar limb as seen from earth), the peak $>$100~MeV photon fluxes, and the speeds of the associated CMEs, coronal waves, and shocks. Note that all flares  were located behind the eastern limb of the Sun.

The Fermi-LAT is an astrophysics observatory in a low Earth orbit and as such it does not have the Sun continuously in its field of view. The coverage for the four BTL flares listed in Table~\ref{tab:flareinfo} was sufficient to perform time-resolved observations of the onset of the events.  Four additional BTL flares have been detected by the LAT over the past 16 years in orbit. They occurred on 2014 Jan 6, 2022 Sep 29, 2024 Feb 14 and 2024 Sep 9. The event of Jan 2014 was only partially observed due to the satellite being in the South Atlantic Anomaly during the onset whereas the event of Sep 2024 \citep[see][]{Gopalswamy2025} did not have sufficient statistics to be able to identify when the flux peaked. The flare of 2022 showed evidence for flare accelerated  particles trapped in a large scale loop~\citep{Pesce-Rollins_2024}. The flare of 2024 Feb 14 was detected by the LAT and a preliminary analysis of the radio and gamma-ray data indicates that the relation found for the four BTL flares reported here, also holds for this event. A in-depth study of this event will be presented in a follow-up work.
The $>$100~MeV gamma-ray data was analyzed via an unbinned likelihood analysis of the Pass 8 Source class events from a 10$^{\circ}$ circular region centered on the Sun and within 100$^{\circ}$ from the local zenith (to reduce contamination from the Earth limb). The LAT flux points reported in Figs.~\ref{fig:lc_20131011} and \ref{fig:lc_20140901} were taken from \cite{2017FermiBTL} and those in Figs.~\ref{fig:lc_20210717} and \ref{fig:lc_20210917} from \cite{Pesce-Rollins_2022}. Further details on the likelihood analysis and the software used can be found in the aforementioned publications.

 Information on the gamma-ray localization for these BTL flares is also reported in the publications cited in the previous paragraph. An updated result for the flare of 2014 Sep 1 can be found in Figure 5 of \cite{Ajello_2021}. However, the 68\% containment error radius is of the order of 100's of arcseconds for all of the BTL flares therefore it is impossible to distinguish any fine structure related to the gamma-ray emission.

Fast CMEs were present in all events. The CME speeds were derived from the CDAW LASCO CME  catalogue \citep{Yashiro2004} by applying a linear fit to the first three data points for each event. This corresponds to a time range that is mostly consistent with the period of observed gamma-ray emission. Note that these are projected speeds since the CME heights are determined in the plane-of-sky.

In all events, large-scale coronal waves were detected in EUV. In SOL20131011 and SOL20140901 the waves were characterized using the 193~\AA\ channel of the Atmospheric Imaging Assembly \citep[AIA;][]{2012SoPh..275...17L} onboard SDO, while the waves in SOL20210717 and SOL20210917 were observed at 195~\AA\ with the Extreme-Ultraviolet Imaging Telescope \citep[EUVI;][]{wuelser2004} onboard STEREO-A. Both channels are sensitive to plasma at around 1.5~MK and generally show the largest response to coronal waves.

To estimate the speed of a wave we define a sector encircling the wave front, assuming that it originates from the flare site and travels along the solar surface. We then produce an intensity–distance profile for the wave front by summing the intensities of the pixels in a selected region in each running-difference frame. We then perform a linear fit to the locations and times of the wave front and derive its propagation speed. Additional details on the intensity–distance profile can be found in~\cite{Pesce-Rollins_2022}.

Dynamic radio spectra are used to search for signatures of coronal shocks. We use data from two stations (Learmonth and San Vito) of the Radio Solar Telescope Network \citep[RSTN;][]{Guidice1981} for three of our events. The RSTN radiospectrographs cover the frequency range of 25--180 MHz with a temporal resolution of 3 sec. For SOL20210717 we employed the Gauribidanur Low-Frequency Solar Spectrograph~\citep{ramesh1998,ramesh2011,kishore2014}, which covers the 40-440 MHz range with a resolution of 0.25 sec. Since type~II burst are generated by plasma emission, we can associate each frequency with an electron density, and with a suitable coronal density model we can estimate heights and speeds. Here, we use a one-fold \cite{Newkirk1961} density model. 
This widely-used model has been shown to be a good representation of densities in the quiet corona \citep[e.g.][]{Warmuth2005}. While type~II bursts can be established still within active regions and can propagate in regions of enhanced densities associated with streamers, they often show nonradial propagation and thus quickly leave the denser regions \citep[e.g.][]{Zucca2018}, so we believe our approach is appropriate. However, we have also computed the shock speeds for a two-fold Newkirk model (for shocks propagating in denser regions) and for the model of \cite{Warmuth2005} which accounts for the transition between active region and quiet coronal densities. The former method resulted in speeds that were higher by 15\%-37\% as compared to the single Newkirk model, while the latter method only resulted in minor changes. Using the different models did not qualitatively affect our results and conclusions.

Note that the speeds derived from radiospectra only reflect the velocity of the source along the density gradient, i.e. mainly along the radial direction in the quiet corona.

\begin{sidewaystable}
\centering
\caption{Event overview. Listed are the event dates, occultation angles of the associated flares, peak $>$100~MeV photon fluxes, and the speeds of the associated CMEs, coronal waves, and shocks (derived from dynamic radio spectra).}
\label{T-simple}
\begin{tabular}{ccccccc}     
\hline							
&	flare & occultation & LAT & early	& EUV & type II	\\
date	& location &  angle & peak flux & CME speed	&	wave speed	 & shock speed	\\
	& &  & [cm$^{-1}$\,s$^{-1}$] & [km\,s$^{-1}$]	&	[km\,s$^{-1}$]	&	[km\,s$^{-1}$]	\\
\hline							
2013 Oct 11	& N21E103 &  12° & $4.9 \pm 0.2\times 10^{-4}$ & $1\,292 \pm 18$	&	$534 \pm 105$	&	$1\,170	\pm 85$ \\
2014 Sep 01	 &  N14E126 & 35° & $6.8 \pm 0.2 \times 10^{-3}$ & $2\,746 \pm 197$	&	$479 \pm 89$	&	$3\,060	\pm 501$ \\
2021 Jul 17	& S20E140 & 46° & $5.1 \pm 1.0 \times 10^{-5}$ & $1\,442 \pm 357$	&	$269 \pm 12$	&	$970 \pm 42$	\\
2021 Sep 17	 & S30E100 & 9° & $7.2 \pm 1.1 \times 10^{-4}$ & $1\,715 \pm 222$	&	$344 \pm 30$ 	&	$2\,880	\pm 85$	\\
\hline							
\end{tabular}
\label{tab:flareinfo}
\end{sidewaystable}

\begin{sidewaysfigure}
    \centering
    \includegraphics[width=\textwidth, trim=0 90 0 90,clip]{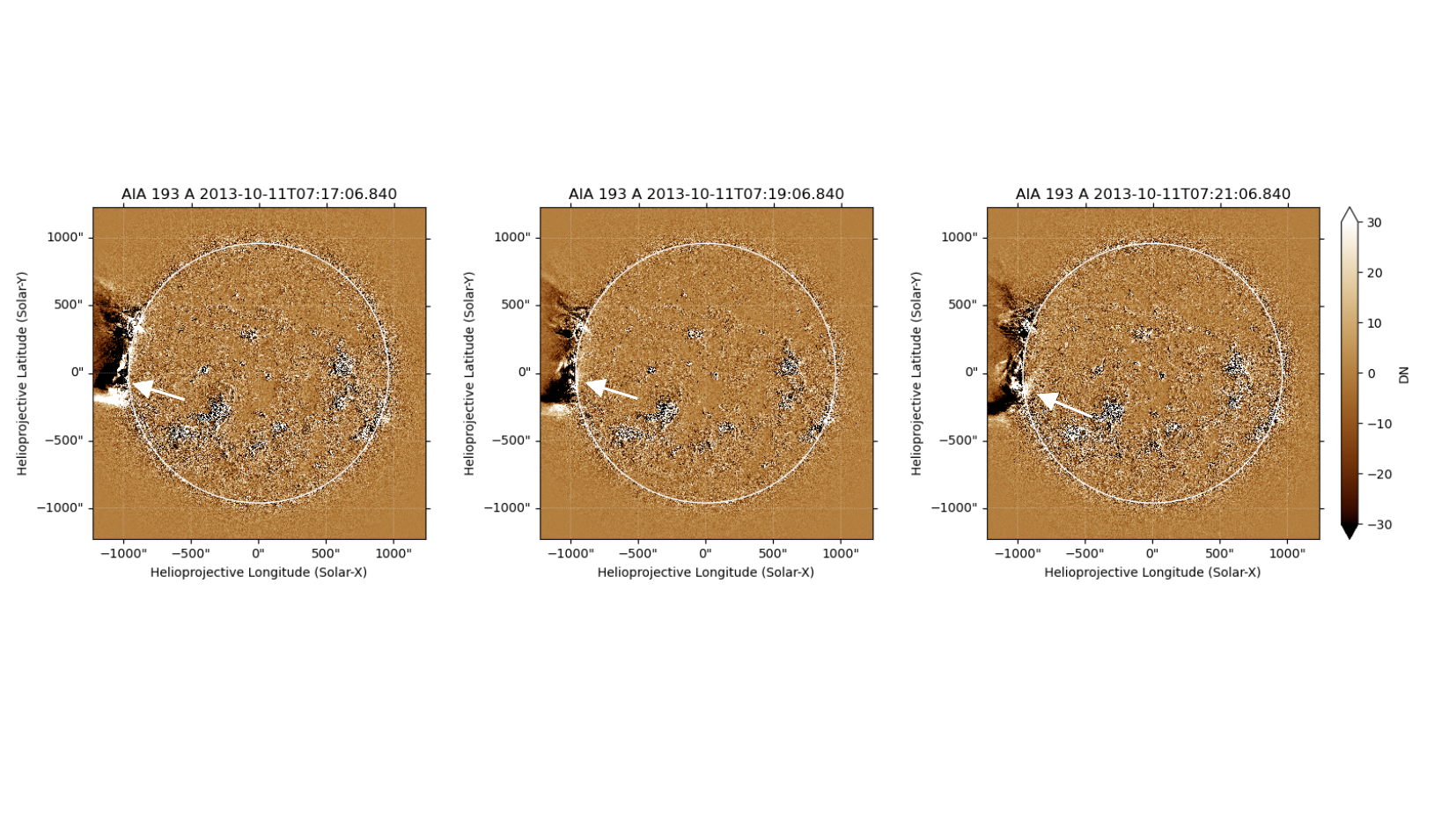}
    \includegraphics[width=\textwidth, trim=0 90 0 90,clip]{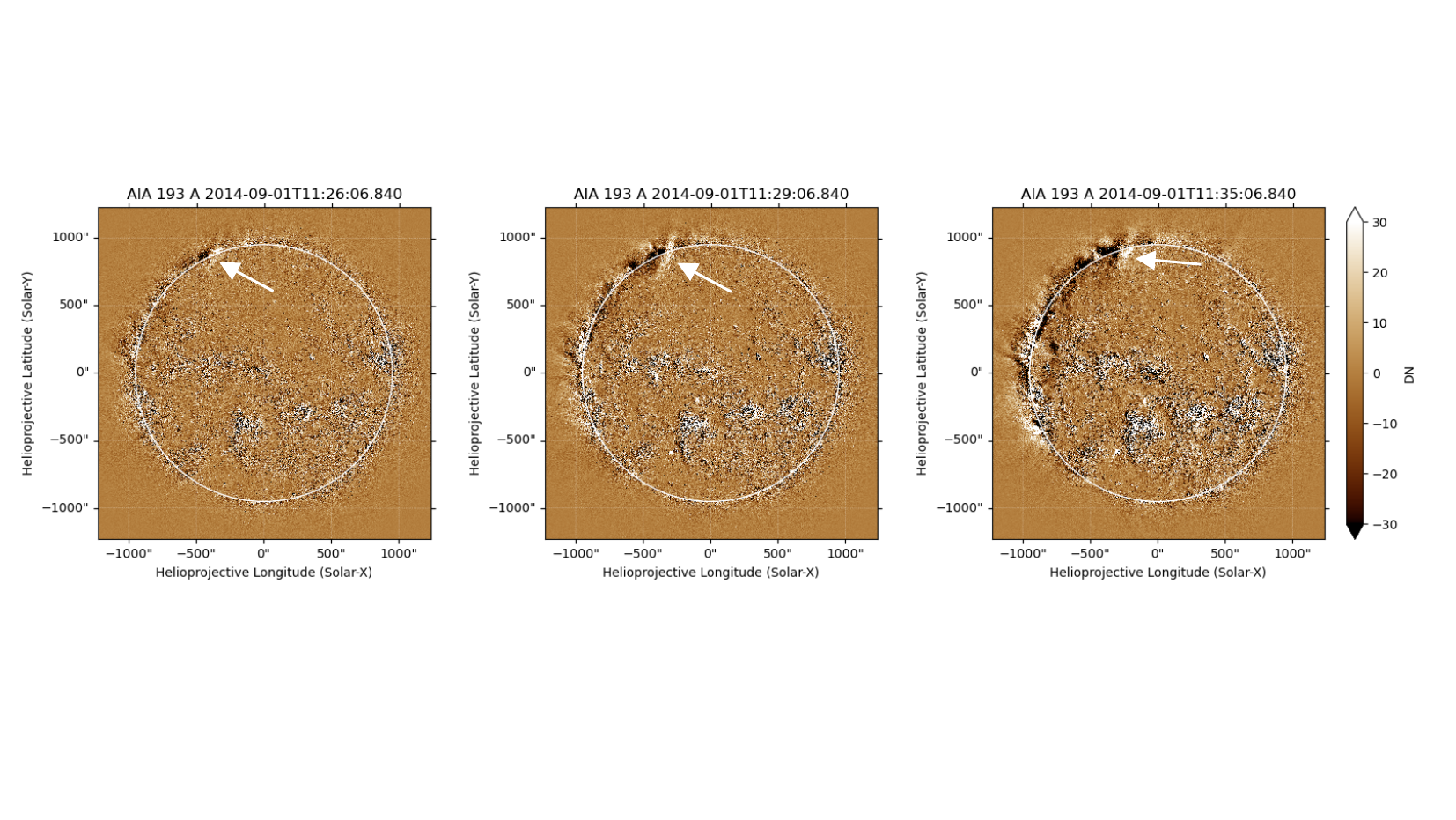}
    \caption{Running-difference AIA 193~\AA\ snapshots of the coronal waves for the BTL flares of 2013 Oct 11 (top panel) and 2014 Sep 1 (bottom panel). The time of the snapshot is reported in each panel. Arrows indicate the position of the coronal wave front.}
    \label{fig:waves}
\end{sidewaysfigure}

\begin{sidewaysfigure}
    \centering
    \includegraphics[width=\textwidth, trim=0 90 0 90,clip]{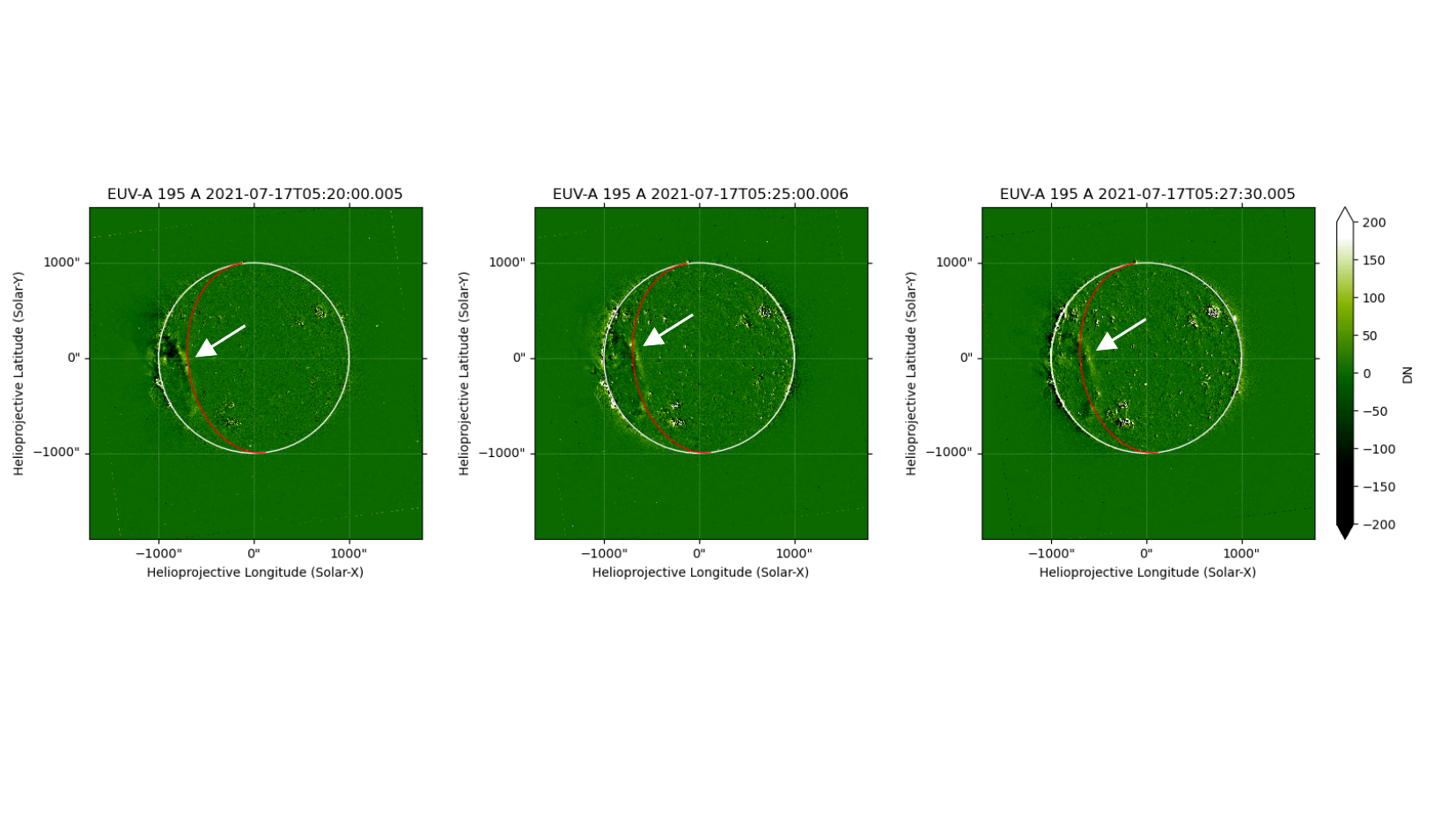}
    \includegraphics[width=\textwidth, trim=0 90 0 90,clip]{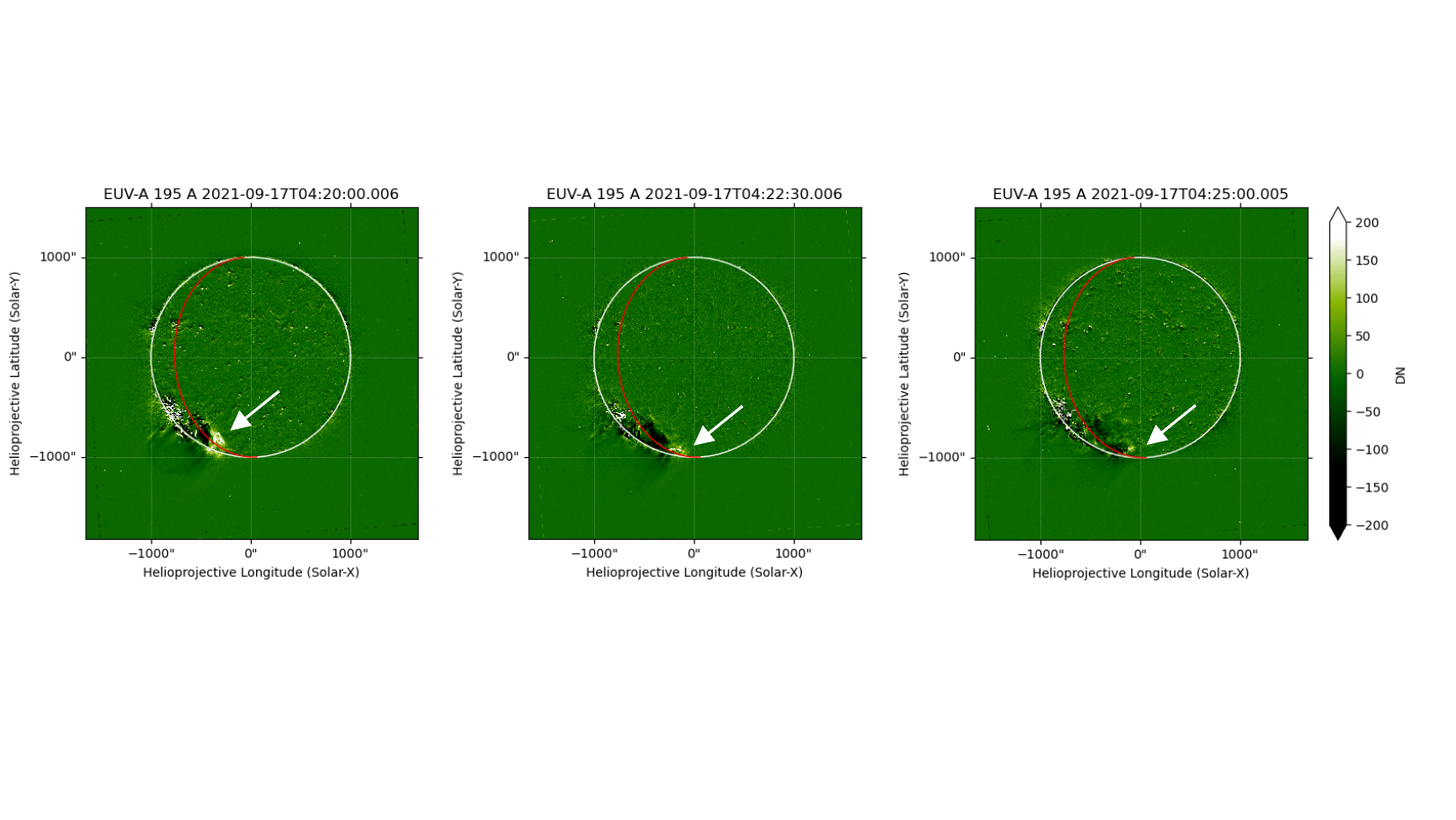}
    \caption{Running-difference EUVI-A 195~\AA\ snapshots of the coronal waves for the BTL flares of 2021 Jul 17 (top panel) and 2021 Sep 17 (bottom panel). The time of the snapshot is reported in each panel. The red line indicates the solar limb as seen from Earth. Arrows indicate the position of the coronal wave front.}
    \label{fig:waves2}
\end{sidewaysfigure}

\section{Results}
\label{sec:res}   
   
Figures~\ref{fig:waves} and \ref{fig:waves2} provide an overview of the propagation of the large-scale coronal EUV waves for all four BTL flares. For each event, we show three running-difference images that illustrate the propagation of the waves, in particular, how they cross onto the hemisphere visible from Earth (the solar limb as seen from Earth is plotted as a red line for the two events where STEREO-A/EUVI data are used). In Figs.~\ref{fig:lc_20131011} to \ref{fig:lc_20210917} we compare the dynamic radio spectra to the gamma-ray lightcurves  as well as to the time evolution of the coronal wave intensities integrated over the hemisphere visible from Earth.

\subsection{2013 Oct 11}  

\begin{figure}    
\centerline{\includegraphics[width=0.9\textwidth,clip=]{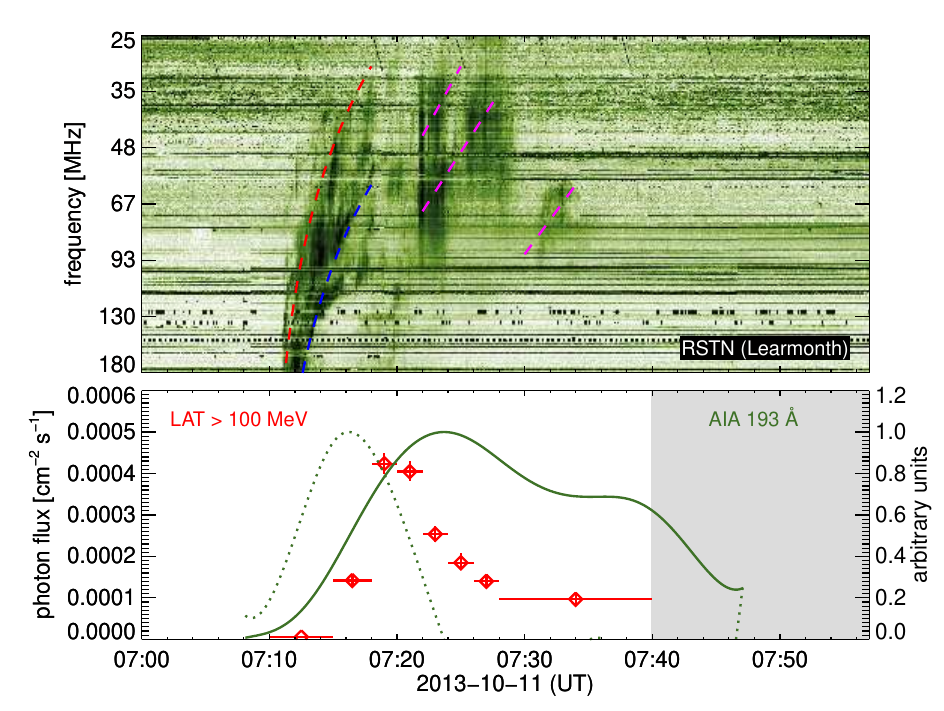}}
\small
\caption{Dynamic radio spectrum and multiwavelength lightcurves of SOL20131011. Top panel: dynamic radiospectrum from the RSTN radiospectrograph at Learmonth showing a type~II radio burst (red dashed line: fundamental emission band; blue line: harmonic band; magenta lines: unspecified bands). Bottom panel: Fermi-LAT $>$100 MeV photon flux (red diamonds) and normalized coronal wave intensity observed at 193~\AA\ with SDO/AIA (green full line; the dashed line shows its derivative). The shaded region in the bottom panel represents Fermi-LAT night.}
\label{fig:lc_20131011}
\end{figure}

The lower panel of Fig.~\ref{fig:lc_20131011} shows the Fermi-LAT gamma-ray photon flux for energies $>$100 MeV (red diamonds) for the BLT flare of 2013 Oct 11 together with the normalized coronal wave intensity observed at 193~\AA\ with SDO/AIA (green full line) and its derivative (green dotted line).  Under the assumption that the disturbance responsible for accelerating the protons is the same that is driving the coronal wave, the time derivative of the intensity enhancement of the coronal wave can be interpreted as providing information on the rate of particles being accelerated—and in turn precipitating to the surface of the Sun. More details on this interpretation can be found in \cite{Pesce-Rollins_2022}. Note that the rise of the gamma-ray flux is consistent with the wave lightcurve. The dynamic radio spectrum in the upper panel shows a metric type~II radioburst that starts at 07:11~UT. The prominent fundamental emission band (which is traced with the red dashed line) starts at a frequency of $\approx 170$~MHz and can be traced to $\approx 35$~MHz. A corresponding harmonic band is seen as well (blue line).  From 07:21~UT on, two other drifting emission lanes become visible, and at 07:30~UT another short-lived drifting burst appears at 90~MHz. We note that this burst is seen during the last Fermi-LAT time bin with significant gamma-ray emission (Fermi-LAT observations stop at 07:40~UT). This means that the complex type~II burst was present during the whole period of gamma-ray emission.

Focusing on the lower frequency edge of the initial fundamental band, which is the most clearly defined feature in this complex event, we find that the radio-emitting shock was formed low in the corona at a height of 70~Mm and could be traced up to a height of nearly one solar radius. The linear shock speed was 1\,170~km\,s$^{-1}$.

\subsection{2014 Sep 01}

\begin{figure}    
\centerline{\includegraphics[width=0.9\textwidth,clip=]{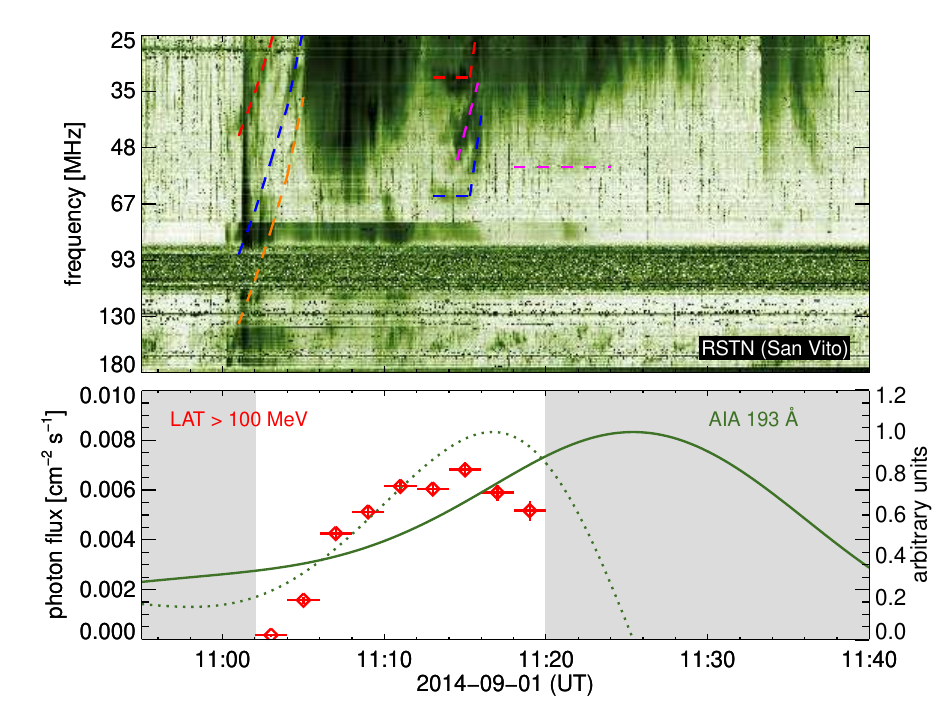}}
\small
\caption{Dynamic radio spectrum and multiwavelength lightcurves of SOL20140901. Top panel: dynamic radio spectrum from the RSTN radiospectrograph at San Vito showing a complex ensemble of radio burst, including several type~II features  (red dashed lines: fundamental emission bands; blue lines: second harmonic bands; orange line: third harmonic band; magenta lines: unspecified bands). Bottom panel: Fermi-LAT $>$100 MeV photon flux (red diamonds) and normalized coronal wave intensity integrated over the hemisphere visible from Earth, observed at 193~\AA\ with SDO/AIA (green full line; the dashed line shows its derivative). The shaded regions in the bottom panel represents Fermi-LAT night.}
\label{fig:lc_20140901}
\end{figure}

\begin{figure}    
\centerline{\includegraphics[width=0.9\textwidth,clip=]{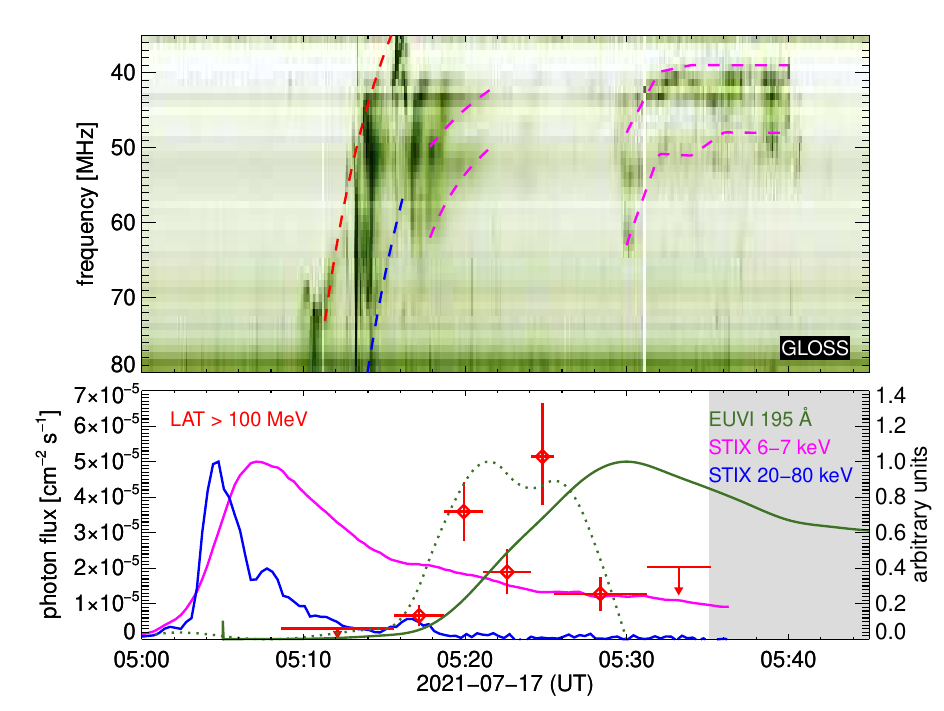}}
\small
\caption{Dynamic radio spectrum and multiwavelength lightcurves of SOL20210717. Top panel: dynamic radio spectrum from the Gauribidanur Low-frequency Solar Spectrograph showing a complex type~II radio burst (red dashed line: fundamental emission band; blue line: harmonic band; magenta lines; unspecified bands). Bottom panel: Fermi-LAT $>$100 MeV photon flux (red diamonds), normalized STIX count rates at 6–7 keV (purple solid line) and 20–80 keV (blue solid line), and normalized coronal wave intensity integrated over the hemisphere visible from Earth, observed at 195~\AA\ with STEREO-A/EUVI (green full line; the dashed line shows its derivative). The shaded region in the bottom panel represents Fermi-LAT night.}
\label{fig:lc_20210717}
\end{figure}

This event shows an extremely complex radio spectrum (see Fig.~\ref{fig:lc_20140901}), aspects of which have been discussed by various authors \citep{Carley2017,Grechnev2018,2020SoPh..295...18G,Kochanov2024}. Here we highlight only a few relevant features. At 11:11~UT, a rapidly drifting type~II burst with fundamental, second and third harmonic band is launched, consistent with the time of the first Fermi-LAT detection. The starting frequency of the fundamental band is 45~MHz, corresponding to a height of 390~Mm. A linear speed of 3\,060~km\,s$^{-1}$ can be derived from the drift rate. This either implies a very fast and strong shock wave, or it could result from the shock crossing a coronal streamer at an oblique angle \citep[][]{Mancuso2002}.

Around 11:14~UT a pair of non-drifting fundamental-harmonic bands is seen for some 2~min, which is interpreted as emission from the flank of a shock that is propagating laterally at a constant height in the corona \citep[see][]{2013NatPh...9..811C,2020A&A...635A..62M,Chrysaphi_2020}. This is immediately followed by a rapid drift to lower frequencies, corresponding to a speed of $\approx$5\,000~km\,s$^{-1}$. This peculiar behaviour could be interpreted in terms of a shock flank that initially propagating laterally at a height of 280~Mm (no radial velocity component implies no frequency drift). When it hits a streamer at an oblique angle, type II radio emission is favored at the intersection point between the streamer the and shock \citep[see][]{Mancuso2002}. As the shock propagates through the streamer, the intersection point rapidly moves upward, which results in a high drift rate. Thus, the derived speed does not correspond to the actual expansion of the shock, but to its intersection with the streamer. Finally, we point out another non-drifting emission band that appears at 11:18~UT at 54~MHz and persists for 6~min, well past the end of Fermi-LAT observations (magenta line in Fig.~\ref{fig:lc_20140901}). Depending on whether the feature is due to fundamental or harmonic emission, the propagation height is 300 or 600~Mm.

\subsection{2021 Jul 17}

Figure~\ref{fig:lc_20210717} shows the lightcurves and the dynamic radio spectrum provided by the GLOSS instrument of the flare of 2021 Jul 17. This event was analyzed in detail in \cite{Pesce-Rollins_2022}, so we provide only a brief discussion here. A type~II burst with fundamental (red line) and harmonic band (blue line) starts at 05:11~UT at 73~MHz (fundamental), coincident in time with the onset of the gamma-ray emission, but clearly delayed from the impulsive HXR peak measured by STIX. After 05:18~UT, two distinct pairs of emission lanes (magenta lines) are seen for which it is unclear whether they are due to fundamental or harmonic emission. The latter of the two pairs is particularly interesting, since after an initial drift it stays at nearly the same frequency. As in the event of 2014 Sep 01, we interpret this in terms of a shock flank that propagates laterally at a constant height. 

From the initial fundamental band, we derive a shock formation height of 245~Mm and a linear speed of 970~km\,s$^{-1}$. For the lower-frequency non-drifting emission band, we determine the heights of 440~Mm and 770~Mm, depending on whether fundamental or harmonic emission is assumed. Since this radio source does not move against the density gradient and we do not have radioheliographic data available, we cannot directly measure its speed. However, if we interpret the band-splitting seen in this burst as being due to emission from up- and downstream of the shock, we can infer the fast-mode magnetosonic Mach number \citep[e.~g.][]{vrsnak2001}. For this event, we obtain Mach numbers between 1.4 and 1.6. We can estimate the magnetosonic speed at the heights corresponding to the radio emission by using the Newkirk density model and the coronal magnetic field strength model by \cite{Dulk1978}. Multiplying this with the Mach numbers, we obtain speeds in the range of 600--800~km\,s$^{-1}$ for the laterally propagating shock, which is not inconsistent with the speed of the earlier drifting burst. Note, however, that the attribution of band-splitting to emission from up- and downstream of the density jump is still being debated \citep[][]{Chrysaphi2018,Bhunia2023}. Even if we accept this interpretation, we have to point out that the derived Mach numbers imply a rather weak shock that may not be able to accelerate ions to high energies. It has been pointed out that electron acceleration that leads to type II radio emission is favored at quasi-perpendicular shocks \citep[e.g.][]{mann2018,Kouloumvakos2021}. This does not preclude the possibility that ions are efficiently accelerated at other locations on the shock where  Mach numbers are higher but the geometry is more oblique \citep[e.g.][]{Rouillard2016}.

\begin{figure}    
\centerline{\includegraphics[width=0.9\textwidth,clip=]{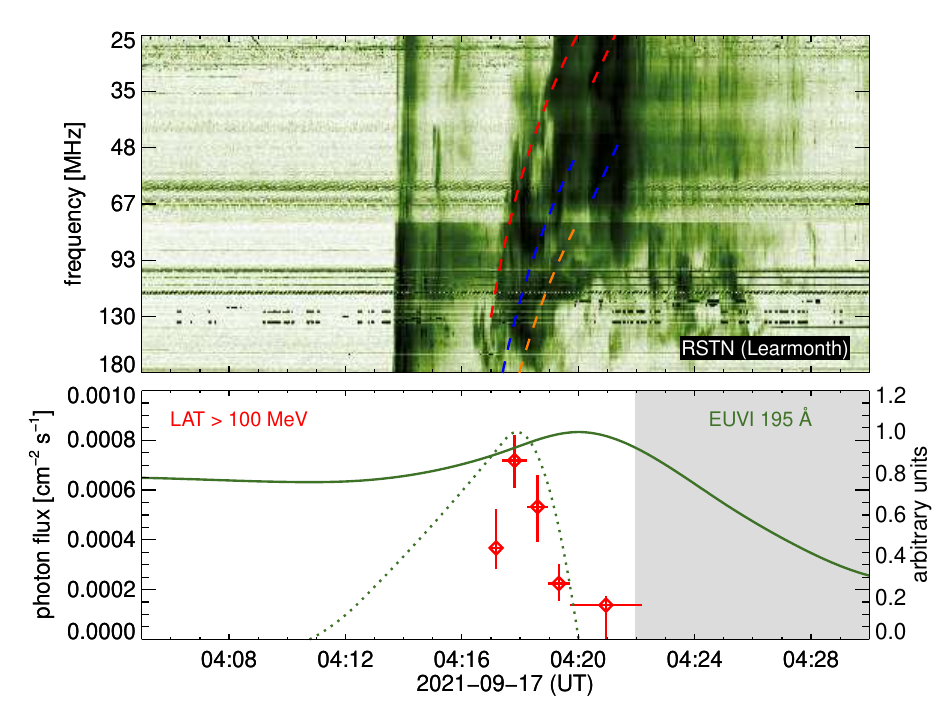}}
\small
\caption{Dynamic radio spectrum and multiwavelength lightcurves of SOL20210917. Top panel: dynamic radio spectrum from the RSTN radiospectrograph at Learmonth showing a rapidly drifting type~II radio burst (red dashed line: fundamental emission band; blue line: harmonic band; orange line: third harmonic band;). Bottom panel: Fermi-LAT $>$100 MeV photon flux (red diamonds) and normalized coronal wave intensity integrated over the hemisphere visible from Earth, observed at 195~\AA\ with STEREO-A/EUVI (green full line; the dashed line shows its derivative). The shaded region in the bottom panel represents Fermi-LAT night.}
\label{fig:lc_20210917}
\end{figure}

\subsection{2021 Sep 17}

Figure~\ref{fig:lc_20210917} shows the lightcurves and the dynamic radio spectrum provided by the Learmonth RSTN station for the flare of 2021 Sep 17. Superposed on the flare continuum emission, a strong, rapidly drifting type~II burst starts at 04:16~UT. Its fundamental band (red dashed line) can be traced from 130~MHz up to 25~MHz, the lower frequency limit of the instrument. A corresponding second harmonic (blue) and a third harmonic band (orange) are visible as well. These bands disappear around 04:20~UT, and after 30~sec another fundamental-harmonic pair appears. This break in the bands implies that the source of radio emission has shifted to another location at the shock .

Note that with a duration of just 5~min, the gamma-ray emission is particularly impulsive in this event, as well as precisely synchronized in time with the type~II burst. The burst onset occurs within the 24-sec time bin of the first LAT detection, and the burst is maintained throughout duration of the gamma-ray emission.

From the initial fundamental band, we derive a shock formation height of 120~Mm and a very high linear shock speed of 2880~km\,s$^{-1}$.

\subsection{Type II burst characteristics}

As a group, the type~II bursts associated with the BTL flares exhibit some specific characteristics compared to a general sample of type~II bursts. The derived shock speeds were in the range of 1\,000--3\,000~km\,s$^{-1}$, which is significantly higher than for typical type~II burst \citep{Robinson1985}, and more consistent with bursts associated with Moreton waves \citep{warmuth2004b,Warmuth2010}, whose generation requires particularly strong shocks in the low corona. Two of our bursts also had high starting frequencies (130 and 170~MHz), which is considerably higher than the typical 80~Mz \citep{vrsnak2001}. This corresponds to low formation heights on the order of 100~Mm. This is consistent with MHD theory which predicts that faster bursts should be formed at lower heights \citep[][]{Vrsnak2000}.

We also note that all events started with a dominating fast-drifting type~II burst which faded after a typical duration of about 10~min. This was followed by more fragmented drifting emission bands. This is consistent with radio emission coming from different parts of the coronal shock. It has been shown that type~II emission is favored for specific shock conditions, e.g. for quasi-perpendicular geometries \citep{Kouloumvakos2021}. For a shock propagating in a complex coronal environment, different locations on the shock will emit type~II radiation at different times, leading to the patchy characteristics of the dynamic radio spectra.  

In several cases also non-drifting emission bands were detected, which is consistent with emission from the flanks of the shock that propagate laterally in the corona. The source heights were in the range of 0.5--1 solar radii. It has been shown that coronal waves propagate faster at larger heights \citep{Hudson2003, veronig2010}, so the larger speeds derived for the drifting bursts and for the non-drifting burst in the 2021~Jul~17 event are not inconsistent with the significantly lower speeds of the associated coronal waves.

\subsection{Correlations}

\begin{figure}    
\begin{center}
\includegraphics[width=0.4\textwidth,clip=]{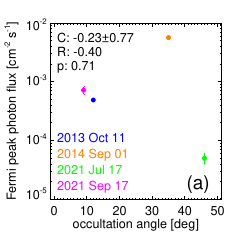}
\includegraphics[width=0.4\textwidth,clip=]{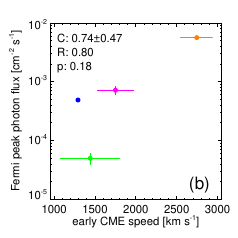}
\end{center}
\vspace{-0.08\textwidth}
\begin{center}
\includegraphics[width=0.4\textwidth,clip=]{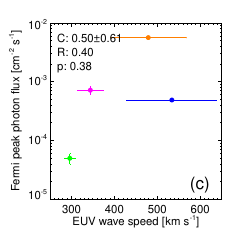}
\includegraphics[width=0.4\textwidth,clip=]{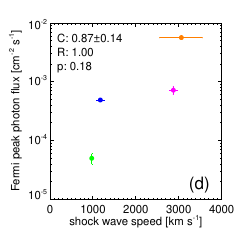}
\end{center}
\vspace{-0.03\textwidth}
\small
\caption{Scatter plots showing the relation between Fermi-LAT peak photon flux versus (a) occultation angle, (b) CME speed, (c) EUV wave speed, and (d) shock speed. Also shown is Pearson's correlation coefficients $C$, Spearman's rank correlation coefficient $R$, and the p-value $p$. Error bars show the one-sigma fit uncertainties. Colors indicate the different events.}
\label{fig:correls}
\end{figure}

If the acceleration of ions to high energies in BTL flares is indeed taking place at coronal shock waves, the gamma-ray parameters should show some correlations with quantities that relate to the shocks. In Fig.~\ref{fig:correls} we show scatter plots of the Fermi-LAT peak photon flux versus four different parameters. Also shown is Pearson's correlation coefficient $C$ (including uncertainties on $C$ based on a bootstrapping approach), Spearman's rank correlation coefficient $R$, and the p-value $p$. Note that we use the logarithm of the peak flux for computing these statistical parameters. In Fig.~\ref{fig:correls}a we plot the peak flux against the occultation angle. Three events are consistent with a decrease of the peak flux for larger occultation angles, which could be interpreted as a weakening of the shock as it propagates to the Earth-facing hemisphere. However, the event of 2014~Sep~01 is a huge outlier. A potential explanation for this could be that the shock flank was exceptionally strong in this event, allowing for a wide extension of a strong shock in the low corona. This appears plausible given the fact that the associated CME was almost twice as fast as in the other events.

With respect to the various speeds (Fig.~\ref{fig:correls}b--d), the LAT peak flux indeed shows a tendency to rise with increasing speed, as would be expected for shock-associated ion acceleration. This tendency appears to be more pronounced for the shock speeds as compared to CMEs or EUV waves. However, the uncertainties on the correlations are high, and the p-values indicate that none of the correlations are statistically significant. 

In addition, the shock speeds we have adopted here were derived from the dominating initial type~II emission bands, which may not accurately represent the speeds of the parts of the shock where the ions are predominantly accelerated. However, at least in the event of 2021~Jul~17 we could show that the speeds corresponding to the initial drifting type~II burst and the later non-drifting burst were quite consistent, as long as the interpretation of band-splitting being due to emission from ahead and behind the shock is accepted.
   
\section{Summary and Conclusion} 
\label{sec:conc} 

We have investigated the relation between high-energy gamma-ray emission and radio-emitting coronal shock waves in four Fermi-LAT behind-the-limb flares. The main results of the present work can be summarized as follows:
\begin{itemize}
\item All BTL gamma-ray flares considered here, were associated with metric type~II radio bursts, which proves that coronal shock waves were present in all events
\item In all events, the starting time of the type~II burst was synchronized with the onset of the gamma-ray emission with an accuracy to within the time bin width of the LAT fluxes (which had a minimum of 24~sec).
\item In all events, the type II radio burst persisted throughout the period of significant gamma-ray emission.
\item The type II bursts associated with the BTL gamma-ray flares had higher speeds and lower formation heights than those of an average sample, which implies the presence of particularly strong shocks.
\item In two events, non-drifting emission bands were seen in addition to the classical drifting type~II features, which indicates that also the flanks of the expanding disturbance that propagate laterally in the corona are at least partly shocked.
\item There is a trend for gamma-ray peak fluxes to increase with rising CME and shock speeds, as would be expected for shock-associated ion acceleration. However, none of these correlations are statistically significant.
\end{itemize}

Going beyond the timing correlations between the gamma-ray emission and the coronal EUV waves reported by \cite{Pesce-Rollins_2022}, these results strongly support the notion that in BTL flares the highly relativistic ions that have access to the Earth-facing hemisphere are accelerated at a large-scale CME-driven coronal shock wave, rather than in the context of a solar flare. Although the type~II radiation is not necessarily emitted by exactly the same regions of the shock that are responsible for ion acceleration, it nevertheless provides important constraints on the shock. This model should be tested in more detail, ideally through data-driven modeling of events for which radioheliographic observations are available.

\begin{acks}
A.W. and S.T. acknowledge funding by the German Space Agency (DLR), grant number \mbox{50 OT 2304}, and by the European Union’s Horizon Europe research and innovation program under grant agreement No. 101134999 (SOLER). This work was performed using the facilities of Institute for Space–Earth Environmental Research (ISEE), Nagoya University. We thank the Gauribidanur team for sharing the GLOSS data. The solar radio facilities in the Gauribidanur observatory are operated by the Indian Institute of Astrophysics. The CDAW CME catalog is generated and maintained at the CDAW Data Center by NASA and The Catholic University of America in cooperation with the Naval Research Laboratory. SOHO is a project of international cooperation between ESA and NASA.
\end{acks}


\bibliographystyle{spr-mp-sola}
\bibliography{references}  

\begin{thebibliography}{63}
\ifx\bisbn     \undefined \def\bisbn  #1{ISBN #1}\fi
\ifx\binits    \undefined \def\binits#1{#1}\fi
\ifx\bauthor   \undefined \def\bauthor#1{#1}\fi
\ifx\batitle   \undefined \def\batitle#1{#1}\fi
\ifx\bjtitle   \undefined \def\bjtitle#1{\textit{#1}}\fi
\ifx\bvolume   \undefined \def\bvolume#1{\textbf{#1}}\fi
\ifx\byear     \undefined \def\byear#1{#1}\fi
\ifx\bissue    \undefined \def\bissue#1{#1}\fi
\ifx\bfpage    \undefined \def\bfpage#1{#1}\fi
\ifx\blpage    \undefined \def\blpage #1{#1}\fi
\ifx\burl      \undefined \def\burl#1{#1}\fi
\ifx\href      \undefined \def\href#1#2{#2}\fi
\ifx\betal     \undefined \def\betal{et al.}\fi
\ifx\bctitle   \undefined \def\bctitle#1{#1}\fi
\ifx\beditor   \undefined \def\beditor#1{#1}\fi
\ifx\bbtitle   \undefined \def\bbtitle#1{\textit{#1}}\fi
\ifx\bedition  \undefined \def\bedition#1{#1}\fi
\ifx\bseriesno \undefined \def\bseriesno#1{\textbf{#1}}\fi
\ifx\blocation \undefined \def\blocation#1{#1}\fi
\ifx\bsertitle \undefined \def\bsertitle#1{\textit{#1}}\fi
\ifx\bsnm      \undefined \def\bsnm#1{#1}\fi
\ifx\bsuffix   \undefined \def\bsuffix#1{#1}\fi
\ifx\bparticle \undefined \def\bparticle#1{#1}\fi
\ifx\barticle  \undefined \def\barticle#1{}\fi
\ifx\binstitute  \undefined \def\binstitute#1{#1}\fi
\ifx\bpublisher  \undefined \def\bpublisher#1{#1}\fi
\ifx\doiurl    \undefined \def\doiurl#1{\href{#1}{DOI}}\fi
\makeatletter
\def\safeHref#1#2#3{\in@{http}{#2}\ifin@\href{#2}{#3}\else\href{#1#2}{#3}\fi}
\makeatother
\ifx\adsurl    \undefined \def\adsurl#1{\safeHref{https://ui.adsabs.harvard.edu/abs/}{#1}{ADS}}\fi
\ifx\arxivurl  \undefined \def\arxivurl#1{\safeHref{http://arxiv.org/abs/}{#1}{arXiv}}\fi
\ifx\botherref \undefined \def\botherref#1{}\fi
\ifx\url       \undefined \def\url#1{#1}\fi
\ifx\bchapter  \undefined \def\bchapter#1{}\fi
\ifx\bbook     \undefined \def\bbook#1{}\fi
\ifx\bcomment  \undefined \def\bcomment#1{#1}\fi
\ifx\oauthor   \undefined \def\oauthor#1{#1}\fi
\ifx\citeauthoryear \undefined\def \citeauthoryear#1{#1}\fi
\def\endbibitem {}
\ifx\bconflocation  \undefined \def\bconflocation#1{#1} \fi

\bibitem[\protect\citeauthoryear{Ackermann et~al.}{2017}]{2017FermiBTL}
\begin{botherref}
\oauthor{\bsnm{Ackermann}, \binits{M.}},
\oauthor{\bsnm{Allafort}, \binits{A.}},
\oauthor{\bsnm{Baldini}, \binits{L.}},
\oauthor{\bsnm{Barbiellini}, \binits{G.}},
\oauthor{\bsnm{Bastieri}, \binits{D.}},
\oauthor{\bsnm{Bellazzini}, \binits{R.}},
\oauthor{\bsnm{Bissaldi}, \binits{E.}},
\oauthor{\bsnm{Bonino}, \binits{R.}},
\oauthor{\bsnm{Bottacini}, \binits{E.}},
\oauthor{\bsnm{Bregeon}, \binits{J.}},
\oauthor{\bsnm{Bruel}, \binits{P.}},
\oauthor{\bsnm{Buehler}, \binits{R.}},
\oauthor{\bsnm{Cameron}, \binits{R.A.}},
\oauthor{\bsnm{Caragiulo}, \binits{M.}},
\oauthor{\bsnm{Caraveo}, \binits{P.A.}},
\oauthor{\bsnm{Cavazzuti}, \binits{E.}},
\oauthor{\bsnm{Cecchi}, \binits{C.}},
\oauthor{\bsnm{Charles}, \binits{E.}},
\oauthor{\bsnm{Ciprini}, \binits{S.}},
\oauthor{\bsnm{Costanza}, \binits{F.}},
\oauthor{\bsnm{Cutini}, \binits{S.}},
\oauthor{\bsnm{D'Ammando}, \binits{F.}},
\oauthor{\bparticle{de} \bsnm{Palma}, \binits{F.}},
\oauthor{\bsnm{Desiante}, \binits{R.}},
\oauthor{\bsnm{Digel}, \binits{S.W.}},
\oauthor{\bsnm{Lalla}, \binits{N.D.}},
\oauthor{\bsnm{Mauro}, \binits{M.D.}},
\oauthor{\bsnm{Venere}, \binits{L.D.}},
\oauthor{\bsnm{Drell}, \binits{P.S.}},
\oauthor{\bsnm{Favuzzi}, \binits{C.}},
\oauthor{\bsnm{Fukazawa}, \binits{Y.}},
\oauthor{\bsnm{Fusco}, \binits{P.}},
\oauthor{\bsnm{Gargano}, \binits{F.}},
\oauthor{\bsnm{Giglietto}, \binits{N.}},
\oauthor{\bsnm{Giordano}, \binits{F.}},
\oauthor{\bsnm{Giroletti}, \binits{M.}},
\oauthor{\bsnm{Grenier}, \binits{I.A.}},
\oauthor{\bsnm{Guillemot}, \binits{L.}},
\oauthor{\bsnm{Guiriec}, \binits{S.}},
\oauthor{\bsnm{Jogler}, \binits{T.}},
\oauthor{\bsnm{J{\'{o}}hannesson}, \binits{G.}},
\oauthor{\bsnm{Kashapova}, \binits{L.}},
\oauthor{\bsnm{Krucker}, \binits{S.}},
\oauthor{\bsnm{Kuss}, \binits{M.}},
\oauthor{\bsnm{Mura}, \binits{G.L.}},
\oauthor{\bsnm{Larsson}, \binits{S.}},
\oauthor{\bsnm{Latronico}, \binits{L.}},
\oauthor{\bsnm{Li}, \binits{J.}},
\oauthor{\bsnm{Liu}, \binits{W.}},
\oauthor{\bsnm{Longo}, \binits{F.}},
\oauthor{\bsnm{Loparco}, \binits{F.}},
\oauthor{\bsnm{Lubrano}, \binits{P.}},
\oauthor{\bsnm{Magill}, \binits{J.D.}},
\oauthor{\bsnm{Maldera}, \binits{S.}},
\oauthor{\bsnm{Manfreda}, \binits{A.}},
\oauthor{\bsnm{Mazziotta}, \binits{M.N.}},
\oauthor{\bsnm{Mitthumsiri}, \binits{W.}},
\oauthor{\bsnm{Mizuno}, \binits{T.}},
\oauthor{\bsnm{Monzani}, \binits{M.E.}},
\oauthor{\bsnm{Morselli}, \binits{A.}},
\oauthor{\bsnm{Moskalenko}, \binits{I.V.}},
\oauthor{\bsnm{Negro}, \binits{M.}},
\oauthor{\bsnm{Nuss}, \binits{E.}},
\oauthor{\bsnm{Ohsugi}, \binits{T.}},
\oauthor{\bsnm{Omodei}, \binits{N.}},
\oauthor{\bsnm{Orlando}, \binits{E.}},
\oauthor{\bsnm{Pal'shin}, \binits{V.}},
\oauthor{\bsnm{Paneque}, \binits{D.}},
\oauthor{\bsnm{Perkins}, \binits{J.S.}},
\oauthor{\bsnm{Pesce-Rollins}, \binits{M.}},
\oauthor{\bsnm{Petrosian}, \binits{V.}},
\oauthor{\bsnm{Piron}, \binits{F.}},
\oauthor{\bsnm{Principe}, \binits{G.}},
\oauthor{\bsnm{Rain{\`{o}}}, \binits{S.}},
\oauthor{\bsnm{Rando}, \binits{R.}},
\oauthor{\bsnm{Razzano}, \binits{M.}},
\oauthor{\bsnm{Reimer}, \binits{O.}},
\oauthor{\bparticle{da} \bsnm{Costa}, \binits{F.R.}},
\oauthor{\bsnm{Sgr{\`{o}}}, \binits{C.}},
\oauthor{\bsnm{Simone}, \binits{D.}},
\oauthor{\bsnm{Siskind}, \binits{E.J.}},
\oauthor{\bsnm{Spada}, \binits{F.}},
\oauthor{\bsnm{Spandre}, \binits{G.}},
\oauthor{\bsnm{Spinelli}, \binits{P.}},
\oauthor{\bsnm{Tajima}, \binits{H.}},
\oauthor{\bsnm{Thayer}, \binits{J.B.}},
\oauthor{\bsnm{Torres}, \binits{D.F.}},
\oauthor{\bsnm{Troja}, \binits{E.}},
\oauthor{\bsnm{Vianello}, \binits{G.}}:
2017,
Fermi-{LAT} Observations of High-energy Behind-the-limb Solar Flares.
\textbf{835},
219.
\doiurl{https://doi.org/10.3847/1538-4357/835/2/219}.
\url{https://doi.org/10.3847/1538-4357/835/2/219}.
\end{botherref}
\endbibitem

\bibitem[\protect\citeauthoryear{Ajello et~al.}{2021a}]{flarecatalog_2021}
\begin{botherref}
\oauthor{\bsnm{Ajello}, \binits{M.}},
\oauthor{\bsnm{Baldini}, \binits{L.}},
\oauthor{\bsnm{Bastieri}, \binits{D.}},
\oauthor{\bsnm{Bellazzini}, \binits{R.}},
\oauthor{\bsnm{Berretta}, \binits{A.}},
\oauthor{\bsnm{Bissaldi}, \binits{E.}},
\oauthor{\bsnm{Blandford}, \binits{R.D.}},
\oauthor{\bsnm{Bonino}, \binits{R.}},
\oauthor{\bsnm{Bruel}, \binits{P.}},
\oauthor{\bsnm{Buson}, \binits{S.}},
\oauthor{\bsnm{Cameron}, \binits{R.A.}},
\oauthor{\bsnm{Caputo}, \binits{R.}},
\oauthor{\bsnm{Cavazzuti}, \binits{E.}},
\oauthor{\bsnm{Cheung}, \binits{C.C.}},
\oauthor{\bsnm{Chiaro}, \binits{G.}},
\oauthor{\bsnm{Costantin}, \binits{D.}},
\oauthor{\bsnm{Cutini}, \binits{S.}},
\oauthor{\bsnm{D'Ammando}, \binits{F.}},
\oauthor{\bparticle{de} \bsnm{Palma}, \binits{F.}},
\oauthor{\bsnm{Desiante}, \binits{R.}},
\oauthor{\bsnm{Lalla}, \binits{N.D.}},
\oauthor{\bsnm{Venere}, \binits{L.D.}},
\oauthor{\bsnm{Dirirsa}, \binits{F.F.}},
\oauthor{\bsnm{Fegan}, \binits{S.J.}},
\oauthor{\bsnm{Fukazawa}, \binits{Y.}},
\oauthor{\bsnm{Funk}, \binits{S.}},
\oauthor{\bsnm{Fusco}, \binits{P.}},
\oauthor{\bsnm{Gargano}, \binits{F.}},
\oauthor{\bsnm{Gasparrini}, \binits{D.}},
\oauthor{\bsnm{Giordano}, \binits{F.}},
\oauthor{\bsnm{Giroletti}, \binits{M.}},
\oauthor{\bsnm{Green}, \binits{D.}},
\oauthor{\bsnm{Guiriec}, \binits{S.}},
\oauthor{\bsnm{Hays}, \binits{E.}},
\oauthor{\bsnm{Hewitt}, \binits{J.W.}},
\oauthor{\bsnm{Horan}, \binits{D.}},
\oauthor{\bsnm{J{\'{o}}hannesson}, \binits{G.}},
\oauthor{\bsnm{Kovac'evic'}, \binits{M.}},
\oauthor{\bsnm{Kuss}, \binits{M.}},
\oauthor{\bsnm{Larsson}, \binits{S.}},
\oauthor{\bsnm{Latronico}, \binits{L.}},
\oauthor{\bsnm{Li}, \binits{J.}},
\oauthor{\bsnm{Longo}, \binits{F.}},
\oauthor{\bsnm{Lovellette}, \binits{M.N.}},
\oauthor{\bsnm{Lubrano}, \binits{P.}},
\oauthor{\bsnm{Maldera}, \binits{S.}},
\oauthor{\bsnm{Manfreda}, \binits{A.}},
\oauthor{\bsnm{Mart{\'{\i}}-Devesa}, \binits{G.}},
\oauthor{\bsnm{Mazziotta}, \binits{M.N.}},
\oauthor{\bsnm{Mereu}, \binits{I.}},
\oauthor{\bsnm{Michelson}, \binits{P.F.}},
\oauthor{\bsnm{Mizuno}, \binits{T.}},
\oauthor{\bsnm{Monzani}, \binits{M.E.}},
\oauthor{\bsnm{Morselli}, \binits{A.}},
\oauthor{\bsnm{Moskalenko}, \binits{I.V.}},
\oauthor{\bsnm{Negro}, \binits{M.}},
\oauthor{\bsnm{Omodei}, \binits{N.}},
\oauthor{\bsnm{Orienti}, \binits{M.}},
\oauthor{\bsnm{Orlando}, \binits{E.}},
\oauthor{\bsnm{Paneque}, \binits{D.}},
\oauthor{\bsnm{Pei}, \binits{Z.}},
\oauthor{\bsnm{Persic}, \binits{M.}},
\oauthor{\bsnm{Pesce-Rollins}, \binits{M.}},
\oauthor{\bsnm{Petrosian}, \binits{V.}},
\oauthor{\bsnm{Piron}, \binits{F.}},
\oauthor{\bsnm{Porter}, \binits{T.A.}},
\oauthor{\bsnm{Principe}, \binits{G.}},
\oauthor{\bsnm{Racusin}, \binits{J.L.}},
\oauthor{\bsnm{Rain{\`{o}}}, \binits{S.}},
\oauthor{\bsnm{Rando}, \binits{R.}},
\oauthor{\bsnm{Rani}, \binits{B.}},
\oauthor{\bsnm{Razzano}, \binits{M.}},
\oauthor{\bsnm{Razzaque}, \binits{S.}},
\oauthor{\bsnm{Reimer}, \binits{A.}},
\oauthor{\bsnm{Reimer}, \binits{O.}},
\oauthor{\bsnm{Serini}, \binits{D.}},
\oauthor{\bsnm{Sgr{\`{o}}}, \binits{C.}},
\oauthor{\bsnm{Siskind}, \binits{E.J.}},
\oauthor{\bsnm{Spandre}, \binits{G.}},
\oauthor{\bsnm{Spinelli}, \binits{P.}},
\oauthor{\bsnm{Tak}, \binits{D.}},
\oauthor{\bsnm{Troja}, \binits{E.}},
\oauthor{\bsnm{Valverde}, \binits{J.}},
\oauthor{\bsnm{Wood}, \binits{K.}},
\oauthor{\bsnm{Zaharijas}, \binits{G.}}:
2021a,
First Fermi-{LAT} Solar Flare Catalog.
\textbf{252},
13.
\doiurl{https://doi.org/10.3847/1538-4365/abd32e}.
\url{https://doi.org/10.3847/1538-4365/abd32e}.
\end{botherref}
\endbibitem

\bibitem[\protect\citeauthoryear{Ajello et~al.}{2021b}]{Ajello_2021}
\begin{barticle}
\bauthor{\bsnm{Ajello}, \binits{M.}},
\bauthor{\bsnm{Baldini}, \binits{L.}},
\bauthor{\bsnm{Bastieri}, \binits{D.}},
\bauthor{\bsnm{Bellazzini}, \binits{R.}},
\bauthor{\bsnm{Berretta}, \binits{A.}},
\bauthor{\bsnm{Bissaldi}, \binits{E.}},
\bauthor{\bsnm{Blandford}, \binits{R.D.}},
\bauthor{\bsnm{Bonino}, \binits{R.}},
\bauthor{\bsnm{Bruel}, \binits{P.}},
\bauthor{\bsnm{Buson}, \binits{S.}},
\bauthor{\bsnm{Cameron}, \binits{R.A.}},
\bauthor{\bsnm{Caputo}, \binits{R.}},
\bauthor{\bsnm{Cavazzuti}, \binits{E.}},
\bauthor{\bsnm{Cheung}, \binits{C.C.}},
\bauthor{\bsnm{Chiaro}, \binits{G.}},
\bauthor{\bsnm{Costantin}, \binits{D.}},
\bauthor{\bsnm{Cutini}, \binits{S.}},
\bauthor{\bsnm{D’Ammando}, \binits{F.}},
\bauthor{\bparticle{de} \bsnm{Palma}, \binits{F.}},
\bauthor{\bsnm{Desiante}, \binits{R.}},
\bauthor{\bsnm{Di~Lalla}, \binits{N.}},
\bauthor{\bsnm{Di~Venere}, \binits{L.}},
\bauthor{\bsnm{Dirirsa}, \binits{F.F.}},
\bauthor{\bsnm{Fegan}, \binits{S.J.}},
\bauthor{\bsnm{Fukazawa}, \binits{Y.}},
\bauthor{\bsnm{Funk}, \binits{S.}},
\bauthor{\bsnm{Fusco}, \binits{P.}},
\bauthor{\bsnm{Gargano}, \binits{F.}},
\bauthor{\bsnm{Gasparrini}, \binits{D.}},
\bauthor{\bsnm{Giordano}, \binits{F.}},
\bauthor{\bsnm{Giroletti}, \binits{M.}},
\bauthor{\bsnm{Green}, \binits{D.}},
\bauthor{\bsnm{Guiriec}, \binits{S.}},
\bauthor{\bsnm{Hays}, \binits{E.}},
\bauthor{\bsnm{Hewitt}, \binits{J.W.}},
\bauthor{\bsnm{Horan}, \binits{D.}},
\bauthor{\bsnm{Jóhannesson}, \binits{G.}},
\bauthor{\bsnm{Kovac’evic’}, \binits{M.}},
\bauthor{\bsnm{Kuss}, \binits{M.}},
\bauthor{\bsnm{Larsson}, \binits{S.}},
\bauthor{\bsnm{Latronico}, \binits{L.}},
\bauthor{\bsnm{Li}, \binits{J.}},
\bauthor{\bsnm{Longo}, \binits{F.}},
\bauthor{\bsnm{Lovellette}, \binits{M.N.}},
\bauthor{\bsnm{Lubrano}, \binits{P.}},
\bauthor{\bsnm{Maldera}, \binits{S.}},
\bauthor{\bsnm{Manfreda}, \binits{A.}},
\bauthor{\bsnm{Martí-Devesa}, \binits{G.}},
\bauthor{\bsnm{Mazziotta}, \binits{M.N.}},
\bauthor{\bsnm{Mereu}, \binits{I.}},
\bauthor{\bsnm{Michelson}, \binits{P.F.}},
\bauthor{\bsnm{Mizuno}, \binits{T.}},
\bauthor{\bsnm{Monzani}, \binits{M.E.}},
\bauthor{\bsnm{Morselli}, \binits{A.}},
\bauthor{\bsnm{Moskalenko}, \binits{I.V.}},
\bauthor{\bsnm{Negro}, \binits{M.}},
\bauthor{\bsnm{Omodei}, \binits{N.}},
\bauthor{\bsnm{Orienti}, \binits{M.}},
\bauthor{\bsnm{Orlando}, \binits{E.}},
\bauthor{\bsnm{Paneque}, \binits{D.}},
\bauthor{\bsnm{Pei}, \binits{Z.}},
\bauthor{\bsnm{Persic}, \binits{M.}},
\bauthor{\bsnm{Pesce-Rollins}, \binits{M.}},
\bauthor{\bsnm{Petrosian}, \binits{V.}},
\bauthor{\bsnm{Piron}, \binits{F.}},
\bauthor{\bsnm{Porter}, \binits{T.A.}},
\bauthor{\bsnm{Principe}, \binits{G.}},
\bauthor{\bsnm{Racusin}, \binits{J.L.}},
\bauthor{\bsnm{Rainò}, \binits{S.}},
\bauthor{\bsnm{Rando}, \binits{R.}},
\bauthor{\bsnm{Rani}, \binits{B.}},
\bauthor{\bsnm{Razzano}, \binits{M.}},
\bauthor{\bsnm{Razzaque}, \binits{S.}},
\bauthor{\bsnm{Reimer}, \binits{A.}},
\bauthor{\bsnm{Reimer}, \binits{O.}},
\bauthor{\bsnm{Serini}, \binits{D.}},
\bauthor{\bsnm{Sgrò}, \binits{C.}},
\bauthor{\bsnm{Siskind}, \binits{E.J.}},
\bauthor{\bsnm{Spandre}, \binits{G.}},
\bauthor{\bsnm{Spinelli}, \binits{P.}},
\bauthor{\bsnm{Tak}, \binits{D.}},
\bauthor{\bsnm{Troja}, \binits{E.}},
\bauthor{\bsnm{Valverde}, \binits{J.}},
\bauthor{\bsnm{Wood}, \binits{K.}},
\bauthor{\bsnm{Zaharijas}, \binits{G.}}:
\byear{2021}b,
\batitle{First Fermi-LAT Solar Flare Catalog}.
\bjtitle{The Astrophysical Journal Supplement Series}
\bvolume{252},
\bfpage{13}.
\doiurl{https://doi.org/10.3847/1538-4365/abd32e}.
\burl{https://dx.doi.org/10.3847/1538-4365/abd32e}.
\end{barticle}
\endbibitem

\bibitem[\protect\citeauthoryear{{Atwood} et~al.}{2009}]{LATPaper}
\begin{barticle}
\bauthor{\bsnm{{Atwood}}, \binits{A.A.} \bsuffix{W.~B.{Abdo}}},
\bauthor{\bsnm{{Ackermann}}, \binits{M.}},
\bauthor{\bsnm{{Ajello}}, \binits{M.}},
\bauthor{\bsnm{{Baldini}}, \binits{L.}},
\bauthor{\bsnm{{Ballet}}, \binits{J.}},
\bauthor{\bsnm{{Barbiellini}}, \binits{G.}},
\bauthor{\bsnm{{Baring}}, \binits{M.G.}},
\bauthor{\bsnm{{Bastieri}}, \binits{D.}},
\bauthor{\bsnm{{Bechtol}}, \binits{K.}},
\bauthor{\bsnm{{Bellazzini}}, \binits{R.}},
\bauthor{\bsnm{{Berenji}}, \binits{B.}},
\bauthor{\bsnm{{Bhat}}, \binits{P.N.}},
\bauthor{\bsnm{{Bissaldi}}, \binits{E.}},
\bauthor{\bsnm{{Blandford}}, \binits{R.D.}},
\bauthor{\bsnm{{Bonamente}}, \binits{E.}},
\bauthor{\bsnm{{Bonnell}}, \binits{J.}},
\bauthor{\bsnm{{Borgland}}, \binits{A.W.}},
\bauthor{\bsnm{{Bouvier}}, \binits{A.}},
\bauthor{\bsnm{{Bregeon}}, \binits{J.}},
\bauthor{\bsnm{{Brigida}}, \binits{M.}},
\bauthor{\bsnm{{Bruel}}, \binits{P.}},
\bauthor{\bsnm{{Buehler}}, \binits{R.}},
\bauthor{\bsnm{{Buson}}, \binits{S.}},
\bauthor{\bsnm{{Caliandro}}, \binits{G.A.}},
\bauthor{\bsnm{{Cameron}}, \binits{R.A.}},
\bauthor{\bsnm{{Caraveo}}, \binits{P.A.}},
\bauthor{\bsnm{{Casandjian}}, \binits{J.M.}},
\bauthor{\bsnm{{Cecchi}}, \binits{C.}},
\bauthor{\bsnm{{Charles}}, \binits{E.}},
\bauthor{\bsnm{{Chekhtman}}, \binits{A.}},
\bauthor{\bsnm{{Chiang}}, \binits{J.}},
\bauthor{\bsnm{{Ciprini}}, \binits{S.}},
\bauthor{\bsnm{{Claus}}, \binits{R.}},
\bauthor{\bsnm{{Connaughton}}, \binits{V.}},
\bauthor{\bsnm{{Conrad}}, \binits{J.}},
\bauthor{\bsnm{{Cutini}}, \binits{S.}},
\bauthor{\bsnm{{de Angelis}}, \binits{A.}},
\bauthor{\bsnm{{de Palma}}, \binits{F.}},
\bauthor{\bsnm{{Dermer}}, \binits{C.D.}},
\bauthor{\bsnm{{Silva}}, \binits{E.d.C.e.}},
\bauthor{\bsnm{{Drell}}, \binits{P.S.}},
\bauthor{\bsnm{{Dubois}}, \binits{R.}},
\bauthor{\bsnm{{Favuzzi}}, \binits{C.}},
\bauthor{\bsnm{{Fukazawa}}, \binits{Y.}},
\bauthor{\bsnm{{Fusco}}, \binits{P.}},
\bauthor{\bsnm{{Gargano}}, \binits{F.}},
\bauthor{\bsnm{{Gehrels}}, \binits{N.}},
\bauthor{\bsnm{{Germani}}, \binits{S.}},
\bauthor{\bsnm{{Giglietto}}, \binits{N.}},
\bauthor{\bsnm{{Giommi}}, \binits{P.}},
\bauthor{\bsnm{{Giordano}}, \binits{F.}},
\bauthor{\bsnm{{Giroletti}}, \binits{M.}},
\bauthor{\bsnm{{Glanzman}}, \binits{T.}},
\bauthor{\bsnm{{ Godfrey}}, \binits{G.}},
\bauthor{\bsnm{{Granot}}, \binits{J.}},
\bauthor{\bsnm{{Grenier}}, \binits{I.A.}},
\bauthor{\bsnm{{Guiriec}}, \binits{S.}},
\bauthor{\bsnm{{Hadasch}}, \binits{D.}},
\bauthor{\bsnm{{Hanabata}}, \binits{Y.}},
\bauthor{\bsnm{{Hughes}}, \binits{R.E.}},
\bauthor{\bsnm{{J{\'o}hannesson}}, \binits{G.}},
\bauthor{\bsnm{{Johnson}}, \binits{A.S.}},
\bauthor{\bsnm{{Kamae}}, \binits{T.}},
\bauthor{\bsnm{{Katagiri}}, \binits{H.}},
\bauthor{\bsnm{{Kataoka}}, \binits{J.}},
\bauthor{\bsnm{{Kerr}}, \binits{M.}},
\bauthor{\bsnm{{Kn{\"o}dlseder}}, \binits{J.}},
\bauthor{\bsnm{{Kuss}}, \binits{M.}},
\bauthor{\bsnm{{Lande}}, \binits{J.}},
\bauthor{\bsnm{{Latronico}}, \binits{L.}},
\bauthor{\bsnm{{Lee}}, \binits{S.-H.}},
\bauthor{\bsnm{{Longo}}, \binits{F.}},
\bauthor{\bsnm{{Loparco}}, \binits{F.}},
\bauthor{\bsnm{{Lott}}, \binits{B.}},
\bauthor{\bsnm{{Lubrano}}, \binits{P.}},
\bauthor{\bsnm{{Mazziotta}}, \binits{M.N.}},
\bauthor{\bsnm{{McEnery}}, \binits{J.E.}},
\bauthor{\bsnm{{M{\'e}sz{\'a}ros}}, \binits{P.}},
\bauthor{\bsnm{{Michelson}}, \binits{P.F.}},
\bauthor{\bsnm{{Mizuno}}, \binits{T.}},
\bauthor{\bsnm{{Moiseev}}, \binits{A.A.}},
\bauthor{\bsnm{{Monzani}}, \binits{M.E.}},
\bauthor{\bsnm{{Morselli}}, \binits{A.}},
\bauthor{\bsnm{{Moskalenko}}, \binits{I.V.}},
\bauthor{\bsnm{{Murgia}}, \binits{S.}},
\bauthor{\bsnm{{Nakamori}}, \binits{T.}},
\bauthor{\bsnm{{Naumann-Godo}}, \binits{M.}},
\bauthor{\bsnm{{Nolan}}, \binits{P.L.}},
\bauthor{\bsnm{{Norris}}, \binits{J.P.}},
\bauthor{\bsnm{{Nuss}}, \binits{E.}},
\bauthor{\bsnm{{Ohsugi}}, \binits{T.}},
\bauthor{\bsnm{{Okumura}}, \binits{A.}},
\bauthor{\bsnm{{Omodei}}, \binits{N.}},
\bauthor{\bsnm{{Orlando}}, \binits{E.}},
\bauthor{\bsnm{{Paciesas}}, \binits{W.S.}},
\bauthor{\bsnm{{Pelassa}}, \binits{V.}},
\bauthor{\bsnm{{Pesce-Rollins}}, \binits{M.}},
\bauthor{\bsnm{{Pierbattista}}, \binits{M.}},
\bauthor{\bsnm{{Piron}}, \binits{F.}},
\bauthor{\bsnm{{Porter}}, \binits{T.A.}},
\bauthor{\bsnm{{Racusin}}, \binits{J.L.}},
\bauthor{\bsnm{{Rain{\`o}}}, \binits{S.}},
\bauthor{\bsnm{{Razzano}}, \binits{M.}},
\bauthor{\bsnm{{Razzaque}}, \binits{S.}},
\bauthor{\bsnm{{Reimer}}, \binits{A.}},
\bauthor{\bsnm{{Reimer}}, \binits{O.}},
\bauthor{\bsnm{{ Reyes}}, \binits{L.C.}},
\bauthor{\bsnm{{Roth}}, \binits{M.}},
\bauthor{\bsnm{{Sadrozinski}}, \binits{H.F.-W.}},
\bauthor{\bsnm{{Sgr{\`o}}}, \binits{C.}},
\bauthor{\bsnm{{Siskind}}, \binits{E.J.}},
\bauthor{\bsnm{{Smith}}, \binits{P.D.}},
\bauthor{\bsnm{{Sonbas}}, \binits{E.}},
\bauthor{\bsnm{{Spandre}}, \binits{G.}},
\bauthor{\bsnm{{Spinelli}}, \binits{P.}},
\bauthor{\bsnm{{Stamatikos}}, \binits{M.}},
\bauthor{\bsnm{{Strickman}}, \binits{M.S.}},
\bauthor{\bsnm{{Takahashi}}, \binits{H.}},
\bauthor{\bsnm{{Tanaka}}, \binits{T.}},
\bauthor{\bsnm{{Tanaka}}, \binits{Y.}},
\bauthor{\bsnm{{Thayer}}, \binits{J.G.}},
\bauthor{\bsnm{{Thayer}}, \binits{J.B.}},
\bauthor{\bsnm{{Torres}}, \binits{D.F.}},
\bauthor{\bsnm{{Tosti}}, \binits{G.}},
\bauthor{\bsnm{{Troja}}, \binits{E.}},
\bauthor{\bsnm{{Uehara}}, \binits{T.}},
\bauthor{\bsnm{{Usher}}, \binits{T.L.}},
\bauthor{\bsnm{{Vandenbroucke}}, \binits{J.}},
\bauthor{\bsnm{{Vasileiou}}, \binits{V.}},
\bauthor{\bsnm{{Vianello}}, \binits{G.}},
\bauthor{\bsnm{{Vilchez}}, \binits{N.}},
\bauthor{\bsnm{{Vitale}}, \binits{V.}},
\bauthor{\bsnm{{von Kienlin}}, \binits{A.}},
\bauthor{\bsnm{{Waite}}, \binits{A.P.}},
\bauthor{\bsnm{{Wang}}, \binits{P.}},
\bauthor{\bsnm{{Winer}}, \binits{B.L.}},
\bauthor{\bsnm{{Wood}}, \binits{K.S.}},
\bauthor{\bsnm{{Yamazaki}}, \binits{R.}},
\bauthor{\bsnm{{Yang}}, \binits{Z.}},
\bauthor{\bsnm{{Ziegler}}, \binits{M.}},
\bauthor{\bsnm{{Piro}}, \binits{L.}},
\bauthor{\bsnm{{Fermi Collaboration}}}:
\byear{2009},
\batitle{{The Large Area Telescope on the Fermi Gamma-Ray Space Telescope Mission}}.
\bjtitle{\apj}
\bvolume{697},
\bfpage{1071}.
\doiurl{https://doi.org/10.1088/0004-637X/697/2/1071}.
\adsurl{2009ApJ...697.1071A}.
\end{barticle}
\endbibitem

\bibitem[\protect\citeauthoryear{{Bhunia} et~al.}{2023}]{Bhunia2023}
\begin{barticle}
\bauthor{\bsnm{{Bhunia}}, \binits{S.}},
\bauthor{\bsnm{{Carley}}, \binits{E.P.}},
\bauthor{\bsnm{{Oberoi}}, \binits{D.}},
\bauthor{\bsnm{{Gallagher}}, \binits{P.T.}}:
\byear{2023},
\batitle{{Imaging-spectroscopy of a band-split type II solar radio burst with the Murchison Widefield Array}}.
\bjtitle{\aap}
\bvolume{670},
\bfpage{A169}.
\doiurl{https://doi.org/10.1051/0004-6361/202244456}.
\adsurl{2023A&A...670A.169B}.
\end{barticle}
\endbibitem

\bibitem[\protect\citeauthoryear{{Carley} et~al.}{2013}]{2013NatPh...9..811C}
\begin{barticle}
\bauthor{\bsnm{{Carley}}, \binits{E.P.}},
\bauthor{\bsnm{{Long}}, \binits{D.M.}},
\bauthor{\bsnm{{Byrne}}, \binits{J.P.}},
\bauthor{\bsnm{{Zucca}}, \binits{P.}},
\bauthor{\bsnm{{Bloomfield}}, \binits{D.S.}},
\bauthor{\bsnm{{McCauley}}, \binits{J.}},
\bauthor{\bsnm{{Gallagher}}, \binits{P.T.}}:
\byear{2013},
\batitle{{Quasiperiodic acceleration of electrons by a plasmoid-driven shock in the solar atmosphere}}.
\bjtitle{Nature Physics}
\bvolume{9},
\bfpage{811}.
\doiurl{https://doi.org/10.1038/nphys2767}.
\adsurl{2013NatPh...9..811C}.
\end{barticle}
\endbibitem

\bibitem[\protect\citeauthoryear{{Carley} et~al.}{2017}]{Carley2017}
\begin{barticle}
\bauthor{\bsnm{{Carley}}, \binits{E.P.}},
\bauthor{\bsnm{{Vilmer}}, \binits{N.}},
\bauthor{\bsnm{{Sim{\~o}es}}, \binits{P.J.A.}},
\bauthor{\bsnm{{{\'O} Fearraigh}}, \binits{B.}}:
\byear{2017},
\batitle{{Estimation of a coronal mass ejection magnetic field strength using radio observations of gyrosynchrotron radiation}}.
\bjtitle{\aap}
\bvolume{608},
\bfpage{A137}.
\doiurl{https://doi.org/10.1051/0004-6361/201731368}.
\adsurl{2017A&A...608A.137C}.
\end{barticle}
\endbibitem

\bibitem[\protect\citeauthoryear{Chrysaphi, Reid, and Kontar}{2020}]{Chrysaphi_2020}
\begin{barticle}
\bauthor{\bsnm{Chrysaphi}, \binits{N.}},
\bauthor{\bsnm{Reid}, \binits{H.A.S.}},
\bauthor{\bsnm{Kontar}, \binits{E.P.}}:
\byear{2020},
\batitle{First Observation of a Type {II} Solar Radio Burst Transitioning between a Stationary and Drifting State}.
\bjtitle{The Astrophysical Journal}
\bvolume{893},
\bfpage{115}.
\doiurl{https://doi.org/10.3847/1538-4357/ab80c1}.
\burl{https://doi.org/10.3847/1538-4357/ab80c1}.
\end{barticle}
\endbibitem

\bibitem[\protect\citeauthoryear{{Chrysaphi} et~al.}{2018}]{Chrysaphi2018}
\begin{barticle}
\bauthor{\bsnm{{Chrysaphi}}, \binits{N.}},
\bauthor{\bsnm{{Kontar}}, \binits{E.P.}},
\bauthor{\bsnm{{Holman}}, \binits{G.D.}},
\bauthor{\bsnm{{Temmer}}, \binits{M.}}:
\byear{2018},
\batitle{{CME-driven Shock and Type II Solar Radio Burst Band Splitting}}.
\bjtitle{\apj}
\bvolume{868},
\bfpage{79}.
\doiurl{https://doi.org/10.3847/1538-4357/aae9e5}.
\adsurl{2018ApJ...868...79C}.
\end{barticle}
\endbibitem

\bibitem[\protect\citeauthoryear{{Cliver}, {Kahler}, and {Vestrand}}{1993}]{cliv93}
\begin{bchapter}
\bauthor{\bsnm{{Cliver}}, \binits{E.W.}},
\bauthor{\bsnm{{Kahler}}, \binits{S.W.}},
\bauthor{\bsnm{{Vestrand}}, \binits{W.T.}}:
\byear{1993},
\bctitle{{On the Origin of Gamma-Ray Emission from the Behind-the-Limb Flare on 29 September 1989}}.
In: \bbtitle{23rd International Cosmic Ray Conference (ICRC23), Volume 3},
\bsertitle{International Cosmic Ray Conference}
\bseriesno{3},
\bfpage{91}.
\adsurl{1993ICRC....3...91C}.
\end{bchapter}
\endbibitem

\bibitem[\protect\citeauthoryear{{Dulk} and {McLean}}{1978}]{Dulk1978}
\begin{barticle}
\bauthor{\bsnm{{Dulk}}, \binits{G.A.}},
\bauthor{\bsnm{{McLean}}, \binits{D.J.}}:
\byear{1978},
\batitle{{Coronal magnetic fields.}}
\bjtitle{\solphys}
\bvolume{57},
\bfpage{279}.
\doiurl{https://doi.org/10.1007/BF00160102}.
\adsurl{1978SoPh...57..279D}.
\end{barticle}
\endbibitem

\bibitem[\protect\citeauthoryear{{Emslie} et~al.}{2012}]{Emslie2012}
\begin{barticle}
\bauthor{\bsnm{{Emslie}}, \binits{A.G.}},
\bauthor{\bsnm{{Dennis}}, \binits{B.R.}},
\bauthor{\bsnm{{Shih}}, \binits{A.Y.}},
\bauthor{\bsnm{{Chamberlin}}, \binits{P.C.}},
\bauthor{\bsnm{{Mewaldt}}, \binits{R.A.}},
\bauthor{\bsnm{{Moore}}, \binits{C.S.}},
\bauthor{\bsnm{{Share}}, \binits{G.H.}},
\bauthor{\bsnm{{Vourlidas}}, \binits{A.}},
\bauthor{\bsnm{{Welsch}}, \binits{B.T.}}:
\byear{2012},
\batitle{{Global Energetics of Thirty-eight Large Solar Eruptive Events}}.
\bjtitle{\apj}
\bvolume{759},
\bfpage{71}.
\doiurl{https://doi.org/10.1088/0004-637X/759/1/71}.
\adsurl{2012ApJ...759...71E}.
\end{barticle}
\endbibitem

\bibitem[\protect\citeauthoryear{{Gopalswamy} et~al.}{2018}]{Gopalswamy2018}
\begin{barticle}
\bauthor{\bsnm{{Gopalswamy}}, \binits{N.}},
\bauthor{\bsnm{{M{\"a}kel{\"a}}}, \binits{P.}},
\bauthor{\bsnm{{Yashiro}}, \binits{S.}},
\bauthor{\bsnm{{Lara}}, \binits{A.}},
\bauthor{\bsnm{{Xie}}, \binits{H.}},
\bauthor{\bsnm{{Akiyama}}, \binits{S.}},
\bauthor{\bsnm{{MacDowall}}, \binits{R.J.}}:
\byear{2018},
\batitle{{Interplanetary Type II Radio Bursts from Wind/WAVES and Sustained Gamma-Ray Emission from Fermi/LAT: Evidence for Shock Source}}.
\bjtitle{\apjl}
\bvolume{868},
\bfpage{L19}.
\doiurl{https://doi.org/10.3847/2041-8213/aaef36}.
\adsurl{2018ApJ...868L..19G}.
\end{barticle}
\endbibitem

\bibitem[\protect\citeauthoryear{Gopalswamy et~al.}{2019}]{Gopalswamy_2019}
\begin{barticle}
\bauthor{\bsnm{Gopalswamy}, \binits{N.}},
\bauthor{\bsnm{Mäkelä}, \binits{P.}},
\bauthor{\bsnm{Yashiro}, \binits{S.}},
\bauthor{\bsnm{Lara}, \binits{A.}},
\bauthor{\bsnm{Akiyama}, \binits{S.}},
\bauthor{\bsnm{Xie}, \binits{H.}}:
\byear{2019},
\batitle{On the Shock Source of Sustained Gamma-Ray Emission from the Sun}.
\bjtitle{Journal of Physics: Conference Series}
\bvolume{1332},
\bfpage{012004}.
\doiurl{https://doi.org/10.1088/1742-6596/1332/1/012004}.
\burl{https://dx.doi.org/10.1088/1742-6596/1332/1/012004}.
\end{barticle}
\endbibitem

\bibitem[\protect\citeauthoryear{{Gopalswamy} et~al.}{2020}]{2020SoPh..295...18G}
\begin{barticle}
\bauthor{\bsnm{{Gopalswamy}}, \binits{N.}},
\bauthor{\bsnm{{M{\"a}kel{\"a}}}, \binits{P.}},
\bauthor{\bsnm{{Yashiro}}, \binits{S.}},
\bauthor{\bsnm{{Akiyama}}, \binits{S.}},
\bauthor{\bsnm{{Xie}}, \binits{H.}},
\bauthor{\bsnm{{Thakur}}, \binits{N.}}:
\byear{2020},
\batitle{{Source of Energetic Protons in the 2014 September 1 Sustained Gamma-ray Emission Event}}.
\bjtitle{\solphys}
\bvolume{295},
\bfpage{18}.
\doiurl{https://doi.org/10.1007/s11207-020-1590-8}.
\adsurl{2020SoPh..295...18G}.
\end{barticle}
\endbibitem

\bibitem[\protect\citeauthoryear{{Gopalswamy} et~al.}{2025}]{Gopalswamy2025}
\begin{botherref}
\oauthor{\bsnm{{Gopalswamy}}, \binits{N.}},
\oauthor{\bsnm{{M{\"a}kel{\"a}}}, \binits{P.}},
\oauthor{\bsnm{{Akiyama}}, \binits{S.}},
\oauthor{\bsnm{{Xie}}, \binits{H.}},
\oauthor{\bsnm{{Yashiro}}, \binits{S.}},
\oauthor{\bsnm{{Bale}}, \binits{S.D.}},
\oauthor{\bsnm{{Wimmer-Schweingruber}}, \binits{R.F.}},
\oauthor{\bsnm{{Kuehl}}, \binits{P.}},
\oauthor{\bsnm{{Krucker}}, \binits{S.}}:
2025,
{Multispacecraft Observations of the 2024 September 9 Backside Solar Eruption that Resulted in a Sustained Gamma Ray Emission Event}.
\textit{arXiv e-prints},
arXiv:2503.23852.
\doiurl{https://doi.org/10.48550/arXiv.2503.23852}.
\adsurl{2025arXiv250323852G}.
\end{botherref}
\endbibitem

\bibitem[\protect\citeauthoryear{{Grechnev} et~al.}{2018}]{Grechnev2018}
\begin{barticle}
\bauthor{\bsnm{{Grechnev}}, \binits{V.V.}},
\bauthor{\bsnm{{Kiselev}}, \binits{V.I.}},
\bauthor{\bsnm{{Kashapova}}, \binits{L.K.}},
\bauthor{\bsnm{{Kochanov}}, \binits{A.A.}},
\bauthor{\bsnm{{Zimovets}}, \binits{I.V.}},
\bauthor{\bsnm{{Uralov}}, \binits{A.M.}},
\bauthor{\bsnm{{Nizamov}}, \binits{B.A.}},
\bauthor{\bsnm{{Grigorieva}}, \binits{I.Y.}},
\bauthor{\bsnm{{Golovin}}, \binits{D.V.}},
\bauthor{\bsnm{{Litvak}}, \binits{M.L.}},
\bauthor{\bsnm{{Mitrofanov}}, \binits{I.G.}},
\bauthor{\bsnm{{Sanin}}, \binits{A.B.}}:
\byear{2018},
\batitle{{Radio, Hard X-Ray, and Gamma-Ray Emissions Associated with a Far-Side Solar Event}}.
\bjtitle{\solphys}
\bvolume{293},
\bfpage{133}.
\doiurl{https://doi.org/10.1007/s11207-018-1352-z}.
\adsurl{2018SoPh..293..133G}.
\end{barticle}
\endbibitem

\bibitem[\protect\citeauthoryear{{Guidice} et~al.}{1981}]{Guidice1981}
\begin{bchapter}
\bauthor{\bsnm{{Guidice}}, \binits{D.A.}},
\bauthor{\bsnm{{Cliver}}, \binits{E.W.}},
\bauthor{\bsnm{{Barron}}, \binits{W.R.}},
\bauthor{\bsnm{{Kahler}}, \binits{S.}}:
\byear{1981},
\bctitle{{The Air Force RSTN System}}.
In: \bbtitle{Bulletin of the American Astronomical Society}
\bseriesno{13},
\bfpage{553}.
\adsurl{1981BAAS...13Q.553G}.
\end{bchapter}
\endbibitem

\bibitem[\protect\citeauthoryear{{Heerikhuisen}, {Litvinenko}, and {Craig}}{2002}]{Her:al-02}
\begin{barticle}
\bauthor{\bsnm{{Heerikhuisen}}, \binits{J.}},
\bauthor{\bsnm{{Litvinenko}}, \binits{Y.E.}},
\bauthor{\bsnm{{Craig}}, \binits{I.J.D.}}:
\byear{2002},
\batitle{{Proton acceleration in analytic reconnecting current sheets}}.
\bjtitle{\apj}
\bvolume{566},
\bfpage{512}.
\doiurl{https://doi.org/10.1086/337957}.
\adsurl{http://cdsads.u-strasbg.fr/abs/2002ApJ...566..512H}.
\end{barticle}
\endbibitem

\bibitem[\protect\citeauthoryear{{Holman} et~al.}{2011}]{Holman2011}
\begin{barticle}
\bauthor{\bsnm{{Holman}}, \binits{G.D.}},
\bauthor{\bsnm{{Aschwanden}}, \binits{M.J.}},
\bauthor{\bsnm{{Aurass}}, \binits{H.}},
\bauthor{\bsnm{{Battaglia}}, \binits{M.}},
\bauthor{\bsnm{{Grigis}}, \binits{P.C.}},
\bauthor{\bsnm{{Kontar}}, \binits{E.P.}},
\bauthor{\bsnm{{Liu}}, \binits{W.}},
\bauthor{\bsnm{{Saint-Hilaire}}, \binits{P.}},
\bauthor{\bsnm{{Zharkova}}, \binits{V.V.}}:
\byear{2011},
\batitle{{Implications of X-ray Observations for Electron Acceleration and Propagation in Solar Flares}}.
\bjtitle{\ssr}
\bvolume{159},
\bfpage{107}.
\doiurl{https://doi.org/10.1007/s11214-010-9680-9}.
\adsurl{2011SSRv..159..107H}.
\end{barticle}
\endbibitem

\bibitem[\protect\citeauthoryear{{Hudson} et~al.}{2003}]{Hudson2003}
\begin{barticle}
\bauthor{\bsnm{{Hudson}}, \binits{H.S.}},
\bauthor{\bsnm{{Khan}}, \binits{J.I.}},
\bauthor{\bsnm{{Lemen}}, \binits{J.R.}},
\bauthor{\bsnm{{Nitta}}, \binits{N.V.}},
\bauthor{\bsnm{{Uchida}}, \binits{Y.}}:
\byear{2003},
\batitle{{Soft X-ray observation of a large-scale coronal wave and its exciter}}.
\bjtitle{\solphys}
\bvolume{212},
\bfpage{121}.
\doiurl{https://doi.org/10.1023/A:1022904125479}.
\adsurl{2003SoPh..212..121H}.
\end{barticle}
\endbibitem

\bibitem[\protect\citeauthoryear{Jin et~al.}{2018}]{Jin_2018}
\begin{barticle}
\bauthor{\bsnm{Jin}, \binits{M.}},
\bauthor{\bsnm{Petrosian}, \binits{V.}},
\bauthor{\bsnm{Liu}, \binits{W.}},
\bauthor{\bsnm{Nitta}, \binits{N.V.}},
\bauthor{\bsnm{Omodei}, \binits{N.}},
\bauthor{\bsnm{Costa}, \binits{F.R.d.}},
\bauthor{\bsnm{Effenberger}, \binits{F.}},
\bauthor{\bsnm{Li}, \binits{G.}},
\bauthor{\bsnm{Pesce-Rollins}, \binits{M.}},
\bauthor{\bsnm{Allafort}, \binits{A.}},
\bauthor{\bsnm{Manchester}, \binits{W.}}:
\byear{2018},
\batitle{Probing the Puzzle of Behind-the-limb gamma-ray Flares: Data-driven Simulations of Magnetic Connectivity and CME-driven Shock Evolution}.
\bjtitle{The Astrophysical Journal}
\bvolume{867},
\bfpage{122}.
\doiurl{https://doi.org/10.3847/1538-4357/aae1fd}.
\burl{https://dx.doi.org/10.3847/1538-4357/aae1fd}.
\end{barticle}
\endbibitem

\bibitem[\protect\citeauthoryear{{Kishore} et~al.}{2014}]{kishore2014}
\begin{barticle}
\bauthor{\bsnm{{Kishore}}, \binits{P.}},
\bauthor{\bsnm{{Kathiravan}}, \binits{C.}},
\bauthor{\bsnm{{Ramesh}}, \binits{R.}},
\bauthor{\bsnm{{Rajalingam}}, \binits{M.}},
\bauthor{\bsnm{{Barve}}, \binits{I.V.}}:
\byear{2014},
\batitle{{Gauribidanur Low-Frequency Solar Spectrograph}}.
\bjtitle{\solphys}
\bvolume{289},
\bfpage{3995}.
\doiurl{https://doi.org/10.1007/s11207-014-0539-1}.
\adsurl{2014SoPh..289.3995K}.
\end{barticle}
\endbibitem

\bibitem[\protect\citeauthoryear{{Kochanov} et~al.}{2024}]{Kochanov2024}
\begin{barticle}
\bauthor{\bsnm{{Kochanov}}, \binits{A.A.}},
\bauthor{\bsnm{{Kiselev}}, \binits{V.I.}},
\bauthor{\bsnm{{Grechnev}}, \binits{V.V.}},
\bauthor{\bsnm{{Uralov}}, \binits{A.M.}}:
\byear{2024},
\batitle{{Localization of the Gamma-Ray Emission Region in the 1 September 2014 Behind-the-Limb Solar Flare According to the Fermi/LAT Data}}.
\bjtitle{\solphys}
\bvolume{299},
\bfpage{18}.
\doiurl{https://doi.org/10.1007/s11207-024-02264-4}.
\adsurl{2024SoPh..299...18K}.
\end{barticle}
\endbibitem

\bibitem[\protect\citeauthoryear{Kouloumvakos et~al.}{2020}]{Kouloumvakos_2020}
\begin{barticle}
\bauthor{\bsnm{Kouloumvakos}, \binits{A.}},
\bauthor{\bsnm{Rouillard}, \binits{A.P.}},
\bauthor{\bsnm{Share}, \binits{G.H.}},
\bauthor{\bsnm{Plotnikov}, \binits{I.}},
\bauthor{\bsnm{Murphy}, \binits{R.}},
\bauthor{\bsnm{Papaioannou}, \binits{A.}},
\bauthor{\bsnm{Wu}, \binits{Y.}}:
\byear{2020},
\batitle{Evidence for a Coronal Shock Wave Origin for Relativistic Protons Producing Solar Gamma-Rays and Observed by Neutron Monitors at Earth}.
\bjtitle{\apj}
\bvolume{893},
\bfpage{76}.
\doiurl{https://doi.org/10.3847/1538-4357/ab8227}.
\burl{https://doi.org/10.3847\%2F1538-4357\%2Fab8227}.
\end{barticle}
\endbibitem

\bibitem[\protect\citeauthoryear{{Kouloumvakos} et~al.}{2021}]{Kouloumvakos2021}
\begin{barticle}
\bauthor{\bsnm{{Kouloumvakos}}, \binits{A.}},
\bauthor{\bsnm{{Rouillard}}, \binits{A.}},
\bauthor{\bsnm{{Warmuth}}, \binits{A.}},
\bauthor{\bsnm{{Magdalenic}}, \binits{J.}},
\bauthor{\bsnm{{Jebaraj}}, \binits{I.C.}},
\bauthor{\bsnm{{Mann}}, \binits{G.}},
\bauthor{\bsnm{{Vainio}}, \binits{R.}},
\bauthor{\bsnm{{Monstein}}, \binits{C.}}:
\byear{2021},
\batitle{{Coronal Conditions for the Occurrence of Type II Radio Bursts}}.
\bjtitle{\apj}
\bvolume{913},
\bfpage{99}.
\doiurl{https://doi.org/10.3847/1538-4357/abf435}.
\adsurl{2021ApJ...913...99K}.
\end{barticle}
\endbibitem

\bibitem[\protect\citeauthoryear{{Krucker} et~al.}{2020}]{2020A&A...642A..15K}
\begin{barticle}
\bauthor{\bsnm{{Krucker}}, \binits{S.}},
\bauthor{\bsnm{{Hurford}}, \binits{G.J.}},
\bauthor{\bsnm{{Grimm}}, \binits{O.}},
\bauthor{\bsnm{{K{\"o}gl}}, \binits{S.}},
\bauthor{\bsnm{{Gr{\"o}belbauer}}, \binits{H.-P.}},
\bauthor{\bsnm{{Etesi}}, \binits{L.}},
\bauthor{\bsnm{{Casadei}}, \binits{D.}},
\bauthor{\bsnm{{Csillaghy}}, \binits{A.}},
\bauthor{\bsnm{{Benz}}, \binits{A.O.}},
\bauthor{\bsnm{{Arnold}}, \binits{N.G.}},
\bauthor{\bsnm{{Molendini}}, \binits{F.}},
\bauthor{\bsnm{{Orleanski}}, \binits{P.}},
\bauthor{\bsnm{{Schori}}, \binits{D.}},
\bauthor{\bsnm{{Xiao}}, \binits{H.}},
\bauthor{\bsnm{{Kuhar}}, \binits{M.}},
\bauthor{\bsnm{{Hochmuth}}, \binits{N.}},
\bauthor{\bsnm{{Felix}}, \binits{S.}},
\bauthor{\bsnm{{Schramka}}, \binits{F.}},
\bauthor{\bsnm{{Marcin}}, \binits{S.}},
\bauthor{\bsnm{{Kobler}}, \binits{S.}},
\bauthor{\bsnm{{Iseli}}, \binits{L.}},
\bauthor{\bsnm{{Dreier}}, \binits{M.}},
\bauthor{\bsnm{{Wiehl}}, \binits{H.J.}},
\bauthor{\bsnm{{Kleint}}, \binits{L.}},
\bauthor{\bsnm{{Battaglia}}, \binits{M.}},
\bauthor{\bsnm{{Lastufka}}, \binits{E.}},
\bauthor{\bsnm{{Sathiapal}}, \binits{H.}},
\bauthor{\bsnm{{Lapadula}}, \binits{K.}},
\bauthor{\bsnm{{Bednarzik}}, \binits{M.}},
\bauthor{\bsnm{{Birrer}}, \binits{G.}},
\bauthor{\bsnm{{Stutz}}, \binits{S.}},
\bauthor{\bsnm{{Wild}}, \binits{C.}},
\bauthor{\bsnm{{Marone}}, \binits{F.}},
\bauthor{\bsnm{{Skup}}, \binits{K.R.}},
\bauthor{\bsnm{{Cichocki}}, \binits{A.}},
\bauthor{\bsnm{{Ber}}, \binits{K.}},
\bauthor{\bsnm{{Rutkowski}}, \binits{K.}},
\bauthor{\bsnm{{Bujwan}}, \binits{W.}},
\bauthor{\bsnm{{Juchnikowski}}, \binits{G.}},
\bauthor{\bsnm{{Winkler}}, \binits{M.}},
\bauthor{\bsnm{{Darmetko}}, \binits{M.}},
\bauthor{\bsnm{{Michalska}}, \binits{M.}},
\bauthor{\bsnm{{Seweryn}}, \binits{K.}},
\bauthor{\bsnm{{Bia{\l}ek}}, \binits{A.}},
\bauthor{\bsnm{{Osica}}, \binits{P.}},
\bauthor{\bsnm{{Sylwester}}, \binits{J.}},
\bauthor{\bsnm{{Kowalinski}}, \binits{M.}},
\bauthor{\bsnm{{{\'S}cis{\l}owski}}, \binits{D.}},
\bauthor{\bsnm{{Siarkowski}}, \binits{M.}},
\bauthor{\bsnm{{St{ke}{\'s}licki}}, \binits{M.}},
\bauthor{\bsnm{{Mrozek}}, \binits{T.}},
\bauthor{\bsnm{{Podg{\'o}rski}}, \binits{P.}},
\bauthor{\bsnm{{Meuris}}, \binits{A.}},
\bauthor{\bsnm{{Limousin}}, \binits{O.}},
\bauthor{\bsnm{{Gevin}}, \binits{O.}},
\bauthor{\bsnm{{Le Mer}}, \binits{I.}},
\bauthor{\bsnm{{Brun}}, \binits{S.}},
\bauthor{\bsnm{{Strugarek}}, \binits{A.}},
\bauthor{\bsnm{{Vilmer}}, \binits{N.}},
\bauthor{\bsnm{{Musset}}, \binits{S.}},
\bauthor{\bsnm{{Maksimovi{\'c}}}, \binits{M.}},
\bauthor{\bsnm{{F{\'a}rn{\'\i}k}}, \binits{F.}},
\bauthor{\bsnm{{Koz{\'a}{\v{c}}ek}}, \binits{Z.}},
\bauthor{\bsnm{{Ka{\v{s}}parov{\'a}}}, \binits{J.}},
\bauthor{\bsnm{{Mann}}, \binits{G.}},
\bauthor{\bsnm{{{\"O}nel}}, \binits{H.}},
\bauthor{\bsnm{{Warmuth}}, \binits{A.}},
\bauthor{\bsnm{{Rendtel}}, \binits{J.}},
\bauthor{\bsnm{{Anderson}}, \binits{J.}},
\bauthor{\bsnm{{Bauer}}, \binits{S.}},
\bauthor{\bsnm{{Dionies}}, \binits{F.}},
\bauthor{\bsnm{{Paschke}}, \binits{J.}},
\bauthor{\bsnm{{Pl{\"u}schke}}, \binits{D.}},
\bauthor{\bsnm{{Woche}}, \binits{M.}},
\bauthor{\bsnm{{Schuller}}, \binits{F.}},
\bauthor{\bsnm{{Veronig}}, \binits{A.M.}},
\bauthor{\bsnm{{Dickson}}, \binits{E.C.M.}},
\bauthor{\bsnm{{Gallagher}}, \binits{P.T.}},
\bauthor{\bsnm{{Maloney}}, \binits{S.A.}},
\bauthor{\bsnm{{Bloomfield}}, \binits{D.S.}},
\bauthor{\bsnm{{Piana}}, \binits{M.}},
\bauthor{\bsnm{{Massone}}, \binits{A.M.}},
\bauthor{\bsnm{{Benvenuto}}, \binits{F.}},
\bauthor{\bsnm{{Massa}}, \binits{P.}},
\bauthor{\bsnm{{Schwartz}}, \binits{R.A.}},
\bauthor{\bsnm{{Dennis}}, \binits{B.R.}},
\bauthor{\bsnm{{van Beek}}, \binits{H.F.}},
\bauthor{\bsnm{{Rodr{\'\i}guez-Pacheco}}, \binits{J.}},
\bauthor{\bsnm{{Lin}}, \binits{R.P.}}:
\byear{2020},
\batitle{{The Spectrometer/Telescope for Imaging X-rays (STIX)}}.
\bjtitle{\aap}
\bvolume{642},
\bfpage{A15}.
\doiurl{https://doi.org/10.1051/0004-6361/201937362}.
\adsurl{2020A&A...642A..15K}.
\end{barticle}
\endbibitem

\bibitem[\protect\citeauthoryear{{Lemen} et~al.}{2012}]{2012SoPh..275...17L}
\begin{barticle}
\bauthor{\bsnm{{Lemen}}, \binits{J.R.}},
\bauthor{\bsnm{{Title}}, \binits{A.M.}},
\bauthor{\bsnm{{Akin}}, \binits{D.J.}},
\bauthor{\bsnm{{Boerner}}, \binits{P.F.}},
\bauthor{\bsnm{{Chou}}, \binits{C.}},
\bauthor{\bsnm{{Drake}}, \binits{J.F.}},
\bauthor{\bsnm{{Duncan}}, \binits{D.W.}},
\bauthor{\bsnm{{Edwards}}, \binits{C.G.}},
\bauthor{\bsnm{{Friedlaender}}, \binits{F.M.}},
\bauthor{\bsnm{{Heyman}}, \binits{G.F.}},
\bauthor{\bsnm{{Hurlburt}}, \binits{N.E.}},
\bauthor{\bsnm{{Katz}}, \binits{N.L.}},
\bauthor{\bsnm{{Kushner}}, \binits{G.D.}},
\bauthor{\bsnm{{Levay}}, \binits{M.}},
\bauthor{\bsnm{{Lindgren}}, \binits{R.W.}},
\bauthor{\bsnm{{Mathur}}, \binits{D.P.}},
\bauthor{\bsnm{{McFeaters}}, \binits{E.L.}},
\bauthor{\bsnm{{Mitchell}}, \binits{S.}},
\bauthor{\bsnm{{Rehse}}, \binits{R.A.}},
\bauthor{\bsnm{{Schrijver}}, \binits{C.J.}},
\bauthor{\bsnm{{Springer}}, \binits{L.A.}},
\bauthor{\bsnm{{Stern}}, \binits{R.A.}},
\bauthor{\bsnm{{Tarbell}}, \binits{T.D.}},
\bauthor{\bsnm{{Wuelser}}, \binits{J.-P.}},
\bauthor{\bsnm{{Wolfson}}, \binits{C.J.}},
\bauthor{\bsnm{{Yanari}}, \binits{C.}},
\bauthor{\bsnm{{Bookbinder}}, \binits{J.A.}},
\bauthor{\bsnm{{Cheimets}}, \binits{P.N.}},
\bauthor{\bsnm{{Caldwell}}, \binits{D.}},
\bauthor{\bsnm{{Deluca}}, \binits{E.E.}},
\bauthor{\bsnm{{Gates}}, \binits{R.}},
\bauthor{\bsnm{{Golub}}, \binits{L.}},
\bauthor{\bsnm{{Park}}, \binits{S.}},
\bauthor{\bsnm{{Podgorski}}, \binits{W.A.}},
\bauthor{\bsnm{{Bush}}, \binits{R.I.}},
\bauthor{\bsnm{{Scherrer}}, \binits{P.H.}},
\bauthor{\bsnm{{Gummin}}, \binits{M.A.}},
\bauthor{\bsnm{{Smith}}, \binits{P.}},
\bauthor{\bsnm{{Auker}}, \binits{G.}},
\bauthor{\bsnm{{Jerram}}, \binits{P.}},
\bauthor{\bsnm{{Pool}}, \binits{P.}},
\bauthor{\bsnm{{Soufli}}, \binits{R.}},
\bauthor{\bsnm{{Windt}}, \binits{D.L.}},
\bauthor{\bsnm{{Beardsley}}, \binits{S.}},
\bauthor{\bsnm{{Clapp}}, \binits{M.}},
\bauthor{\bsnm{{Lang}}, \binits{J.}},
\bauthor{\bsnm{{Waltham}}, \binits{N.}}:
\byear{2012},
\batitle{{The Atmospheric Imaging Assembly (AIA) on the Solar Dynamics Observatory (SDO)}}.
\bjtitle{\solphys}
\bvolume{275},
\bfpage{17}.
\doiurl{https://doi.org/10.1007/s11207-011-9776-8}.
\adsurl{2012SoPh..275...17L}.
\end{barticle}
\endbibitem

\bibitem[\protect\citeauthoryear{{Lin} et~al.}{2003}]{Lin2003}
\begin{barticle}
\bauthor{\bsnm{{Lin}}, \binits{R.P.}},
\bauthor{\bsnm{{Krucker}}, \binits{S.}},
\bauthor{\bsnm{{Hurford}}, \binits{G.J.}},
\bauthor{\bsnm{{Smith}}, \binits{D.M.}},
\bauthor{\bsnm{{Hudson}}, \binits{H.S.}},
\bauthor{\bsnm{{Holman}}, \binits{G.D.}},
\bauthor{\bsnm{{Schwartz}}, \binits{R.A.}},
\bauthor{\bsnm{{Dennis}}, \binits{B.R.}},
\bauthor{\bsnm{{Share}}, \binits{G.H.}},
\bauthor{\bsnm{{Murphy}}, \binits{R.J.}},
\bauthor{\bsnm{{Emslie}}, \binits{A.G.}},
\bauthor{\bsnm{{Johns-Krull}}, \binits{C.}},
\bauthor{\bsnm{{Vilmer}}, \binits{N.}}:
\byear{2003},
\batitle{{RHESSI Observations of Particle Acceleration and Energy Release in an Intense Solar Gamma-Ray Line Flare}}.
\bjtitle{\apjl}
\bvolume{595},
\bfpage{L69}.
\doiurl{https://doi.org/10.1086/378932}.
\adsurl{2003ApJ...595L..69L}.
\end{barticle}
\endbibitem

\bibitem[\protect\citeauthoryear{{Litvinenko} and {Somov}}{1995}]{Lit:Som-95}
\begin{barticle}
\bauthor{\bsnm{{Litvinenko}}, \binits{Y.E.}},
\bauthor{\bsnm{{Somov}}, \binits{B.V.}}:
\byear{1995},
\batitle{{Relativistic acceleration of protons in reconnecting current sheets of solar flares}}.
\bjtitle{\solphys}
\bvolume{158},
\bfpage{317}.
\end{barticle}
\endbibitem

\bibitem[\protect\citeauthoryear{{Mancuso} et~al.}{2002}]{Mancuso2002}
\begin{barticle}
\bauthor{\bsnm{{Mancuso}}, \binits{S.}},
\bauthor{\bsnm{{Raymond}}, \binits{J.C.}},
\bauthor{\bsnm{{Kohl}}, \binits{J.}},
\bauthor{\bsnm{{Ko}}, \binits{Y.-K.}},
\bauthor{\bsnm{{Uzzo}}, \binits{M.}},
\bauthor{\bsnm{{Wu}}, \binits{R.}}:
\byear{2002},
\batitle{{UVCS/SOHO observations of a CME-driven shock: Consequences on ion heating mechanisms behind a coronal shock}}.
\bjtitle{\aap}
\bvolume{383},
\bfpage{267}.
\doiurl{https://doi.org/10.1051/0004-6361:20011721}.
\adsurl{2002A&A...383..267M}.
\end{barticle}
\endbibitem

\bibitem[\protect\citeauthoryear{{Mandzhavidze} and {Ramaty}}{1992}]{1992ApJ...396L.111M}
\begin{barticle}
\bauthor{\bsnm{{Mandzhavidze}}, \binits{N.}},
\bauthor{\bsnm{{Ramaty}}, \binits{R.}}:
\byear{1992},
\batitle{{Gamma Rays from Pion Decay: Evidence for Long-Term Trappings of Particles in Solar Flares}}.
\bjtitle{\apjl}
\bvolume{396},
\bfpage{L111}.
\doiurl{https://doi.org/10.1086/186529}.
\adsurl{1992ApJ...396L.111M}.
\end{barticle}
\endbibitem

\bibitem[\protect\citeauthoryear{{Mann} et~al.}{2018}]{mann2018}
\begin{barticle}
\bauthor{\bsnm{{Mann}}, \binits{G.}},
\bauthor{\bsnm{{Melnik}}, \binits{V.N.}},
\bauthor{\bsnm{{Rucker}}, \binits{H.O.}},
\bauthor{\bsnm{{Konovalenko}}, \binits{A.A.}},
\bauthor{\bsnm{{Brazhenko}}, \binits{A.I.}}:
\byear{2018},
\batitle{{Radio signatures of shock-accelerated electron beams in the solar corona}}.
\bjtitle{\aap}
\bvolume{609},
\bfpage{A41}.
\doiurl{https://doi.org/10.1051/0004-6361/201730546}.
\adsurl{2018A&A...609A..41M}.
\end{barticle}
\endbibitem

\bibitem[\protect\citeauthoryear{{Morosan} et~al.}{2020}]{2020A&A...635A..62M}
\begin{barticle}
\bauthor{\bsnm{{Morosan}}, \binits{D.E.}},
\bauthor{\bsnm{{Palmerio}}, \binits{E.}},
\bauthor{\bsnm{{Pomoell}}, \binits{J.}},
\bauthor{\bsnm{{Vainio}}, \binits{R.}},
\bauthor{\bsnm{{Palmroth}}, \binits{M.}},
\bauthor{\bsnm{{Kilpua}}, \binits{E.K.J.}}:
\byear{2020},
\batitle{{Three-dimensional reconstruction of multiple particle acceleration regions during a coronal mass ejection}}.
\bjtitle{\aap}
\bvolume{635},
\bfpage{A62}.
\doiurl{https://doi.org/10.1051/0004-6361/201937133}.
\adsurl{2020A&A...635A..62M}.
\end{barticle}
\endbibitem

\bibitem[\protect\citeauthoryear{{Muhr} et~al.}{2014}]{muhr2014}
\begin{barticle}
\bauthor{\bsnm{{Muhr}}, \binits{N.}},
\bauthor{\bsnm{{Veronig}}, \binits{A.M.}},
\bauthor{\bsnm{{Kienreich}}, \binits{I.W.}},
\bauthor{\bsnm{{Vr{\v{s}}nak}}, \binits{B.}},
\bauthor{\bsnm{{Temmer}}, \binits{M.}},
\bauthor{\bsnm{{Bein}}, \binits{B.M.}}:
\byear{2014},
\batitle{{Statistical Analysis of Large-Scale EUV Waves Observed by STEREO/EUVI}}.
\bjtitle{\solphys}
\bvolume{289},
\bfpage{4563}.
\doiurl{https://doi.org/10.1007/s11207-014-0594-7}.
\adsurl{2014SoPh..289.4563M}.
\end{barticle}
\endbibitem

\bibitem[\protect\citeauthoryear{{Newkirk}}{1961}]{Newkirk1961}
\begin{barticle}
\bauthor{\bsnm{{Newkirk}}, \binits{G.} \bsuffix{Jr.}}:
\byear{1961},
\batitle{{The Solar Corona in Active Regions and the Thermal Origin of the Slowly Varying Component of Solar Radio Radiation.}}
\bjtitle{\apj}
\bvolume{133},
\bfpage{983}.
\doiurl{https://doi.org/10.1086/147104}.
\adsurl{1961ApJ...133..983N}.
\end{barticle}
\endbibitem

\bibitem[\protect\citeauthoryear{Pesce-Rollins et~al.}{2022}]{Pesce-Rollins_2022}
\begin{barticle}
\bauthor{\bsnm{Pesce-Rollins}, \binits{M.}},
\bauthor{\bsnm{Omodei}, \binits{N.}},
\bauthor{\bsnm{Krucker}, \binits{S.}},
\bauthor{\bsnm{Di~Lalla}, \binits{N.}},
\bauthor{\bsnm{Wang}, \binits{W.}},
\bauthor{\bsnm{Battaglia}, \binits{A.F.}},
\bauthor{\bsnm{Warmuth}, \binits{A.}},
\bauthor{\bsnm{Veronig}, \binits{A.M.}},
\bauthor{\bsnm{Baldini}, \binits{L.}}:
\byear{2022},
\batitle{The Coupling of an EUV Coronal Wave and Ion Acceleration in a Fermi-LAT Behind-the-Limb Solar Flare}.
\bjtitle{The Astrophysical Journal}
\bvolume{929},
\bfpage{172}.
\doiurl{https://doi.org/10.3847/1538-4357/ac5f0c}.
\burl{https://dx.doi.org/10.3847/1538-4357/ac5f0c}.
\end{barticle}
\endbibitem

\bibitem[\protect\citeauthoryear{{Pesce-Rollins} et~al.}{2024}]{Pesce-Rollins_2024}
\begin{barticle}
\bauthor{\bsnm{{Pesce-Rollins}}, \binits{M.}},
\bauthor{\bsnm{{Klein, Karl-Ludwig}}},
\bauthor{\bsnm{{Krucker, Säm}}},
\bauthor{\bsnm{{Warmuth, Alexander}}},
\bauthor{\bsnm{{Veronig, Astrid M.}}},
\bauthor{\bsnm{{Omodei, Nicola}}},
\bauthor{\bsnm{{Monstein, Christian}}}:
\byear{2024},
\batitle{Evidence for flare-accelerated particles in large scale loops in the behind-the-limb gamma-ray solar flare of September 29, 2022}.
\bjtitle{AA}
\bvolume{683},
\bfpage{A208}.
\doiurl{https://doi.org/10.1051/0004-6361/202348088}.
\burl{https://doi.org/10.1051/0004-6361/202348088}.
\end{barticle}
\endbibitem

\bibitem[\protect\citeauthoryear{{Plotnikov}, {Rouillard}, and {Share}}{2017}]{Plotnikov_2017}
\begin{barticle}
\bauthor{\bsnm{{Plotnikov}}, \binits{I.}},
\bauthor{\bsnm{{Rouillard}}, \binits{A.P.}},
\bauthor{\bsnm{{Share}}, \binits{G.H.}}:
\byear{2017},
\batitle{The magnetic connectivity of coronal shocks from behind-the-limb flares to the visible solar surface during events}.
\bjtitle{A\&A}
\bvolume{608},
\bfpage{A43}.
\doiurl{https://doi.org/10.1051/0004-6361/201730804}.
\burl{https://doi.org/10.1051/0004-6361/201730804}.
\end{barticle}
\endbibitem

\bibitem[\protect\citeauthoryear{{Ramesh}}{2011}]{ramesh2011}
\begin{bchapter}
\bauthor{\bsnm{{Ramesh}}, \binits{R.}}:
\byear{2011},
\bctitle{{Low frequency solar radio astronomy at the Indian Institute of Astrophysics (IIA)}}.
In: \bbtitle{???},
\bsertitle{Astronomical Society of India Conference Series}
\bseriesno{2},
\bfpage{55}.
\adsurl{2011ASInC...2...55R}.
\end{bchapter}
\endbibitem

\bibitem[\protect\citeauthoryear{{Ramesh} et~al.}{1998}]{ramesh1998}
\begin{barticle}
\bauthor{\bsnm{{Ramesh}}, \binits{R.}},
\bauthor{\bsnm{{Subramanian}}, \binits{K.R.}},
\bauthor{\bsnm{{Sundararajan}}, \binits{M.S.}},
\bauthor{\bsnm{{Sastry}}, \binits{C.V.}}:
\byear{1998},
\batitle{{The Gauribidanur Radioheliograph}}.
\bjtitle{\solphys}
\bvolume{181},
\bfpage{439}.
\doiurl{https://doi.org/10.1023/A:1005075003370}.
\adsurl{1998SoPh..181..439R}.
\end{barticle}
\endbibitem

\bibitem[\protect\citeauthoryear{{Rank} et~al.}{2001}]{rank2001}
\begin{barticle}
\bauthor{\bsnm{{Rank}}, \binits{G.}},
\bauthor{\bsnm{{Ryan}}, \binits{J.}},
\bauthor{\bsnm{{Debrunner, H.}}},
\bauthor{\bsnm{{McConnell, M.}}},
\bauthor{\bsnm{{Sch\"onfelder, V.}}}:
\byear{2001},
\batitle{Extended gamma-ray emission of the solar flares in june 1991}.
\bjtitle{A\&A}
\bvolume{378},
\bfpage{1046}.
\doiurl{https://doi.org/10.1051/0004-6361:20011060}.
\burl{https://doi.org/10.1051/0004-6361:20011060}.
\end{barticle}
\endbibitem

\bibitem[\protect\citeauthoryear{{Robinson}}{1985}]{Robinson1985}
\begin{barticle}
\bauthor{\bsnm{{Robinson}}, \binits{R.D.}}:
\byear{1985},
\batitle{{Velocities of Type-II Solar Radio Events}}.
\bjtitle{\solphys}
\bvolume{95},
\bfpage{343}.
\doiurl{https://doi.org/10.1007/BF00152411}.
\adsurl{1985SoPh...95..343R}.
\end{barticle}
\endbibitem

\bibitem[\protect\citeauthoryear{{Rouillard} et~al.}{2016}]{Rouillard2016}
\begin{barticle}
\bauthor{\bsnm{{Rouillard}}, \binits{A.P.}},
\bauthor{\bsnm{{Plotnikov}}, \binits{I.}},
\bauthor{\bsnm{{Pinto}}, \binits{R.F.}},
\bauthor{\bsnm{{Tirole}}, \binits{M.}},
\bauthor{\bsnm{{Lavarra}}, \binits{M.}},
\bauthor{\bsnm{{Zucca}}, \binits{P.}},
\bauthor{\bsnm{{Vainio}}, \binits{R.}},
\bauthor{\bsnm{{Tylka}}, \binits{A.J.}},
\bauthor{\bsnm{{Vourlidas}}, \binits{A.}},
\bauthor{\bsnm{{De Rosa}}, \binits{M.L.}},
\bauthor{\bsnm{{Linker}}, \binits{J.}},
\bauthor{\bsnm{{Warmuth}}, \binits{A.}},
\bauthor{\bsnm{{Mann}}, \binits{G.}},
\bauthor{\bsnm{{Cohen}}, \binits{C.M.S.}},
\bauthor{\bsnm{{Mewaldt}}, \binits{R.A.}}:
\byear{2016},
\batitle{{Deriving the Properties of Coronal Pressure Fronts in 3D: Application to the 2012 May 17 Ground Level Enhancement}}.
\bjtitle{\apj}
\bvolume{833},
\bfpage{45}.
\doiurl{https://doi.org/10.3847/1538-4357/833/1/45}.
\adsurl{2016ApJ...833...45R}.
\end{barticle}
\endbibitem

\bibitem[\protect\citeauthoryear{{Ryan}}{2000}]{ryan00}
\begin{barticle}
\bauthor{\bsnm{{Ryan}}, \binits{J.M.}}:
\byear{2000},
\batitle{{Long-Duration Solar Gamma-Ray Flares}}.
\bjtitle{\ssr}
\bvolume{93},
\bfpage{581}.
\adsurl{2000SSRv...93..581R}.
\end{barticle}
\endbibitem

\bibitem[\protect\citeauthoryear{{Ryan} and {Lee}}{1991}]{ryanlee91}
\begin{barticle}
\bauthor{\bsnm{{Ryan}}, \binits{J.M.}},
\bauthor{\bsnm{{Lee}}, \binits{M.A.}}:
\byear{1991},
\batitle{{On the Transport and Acceleration of Solar Flare Particles in a Coronal Loop}}.
\bjtitle{\apj}
\bvolume{368},
\bfpage{316}.
\doiurl{https://doi.org/10.1086/169695}.
\adsurl{1991ApJ...368..316R}.
\end{barticle}
\endbibitem

\bibitem[\protect\citeauthoryear{Share et~al.}{2018}]{Share_2018}
\begin{barticle}
\bauthor{\bsnm{Share}, \binits{G.H.}},
\bauthor{\bsnm{Murphy}, \binits{R.J.}},
\bauthor{\bsnm{White}, \binits{S.M.}},
\bauthor{\bsnm{Tolbert}, \binits{A.K.}},
\bauthor{\bsnm{Dennis}, \binits{B.R.}},
\bauthor{\bsnm{Schwartz}, \binits{R.A.}},
\bauthor{\bsnm{Smart}, \binits{D.F.}},
\bauthor{\bsnm{Shea}, \binits{M.A.}}:
\byear{2018},
\batitle{Characteristics of Late-phase greater than 100 MeV Gamma-Ray Emission in Solar Eruptive Events}.
\bjtitle{The Astrophysical Journal}
\bvolume{869},
\bfpage{182}.
\doiurl{https://doi.org/10.3847/1538-4357/aaebf7}.
\burl{https://dx.doi.org/10.3847/1538-4357/aaebf7}.
\end{barticle}
\endbibitem

\bibitem[\protect\citeauthoryear{{Shih}, {Lin}, and {Smith}}{2009}]{Shi2009}
\begin{barticle}
\bauthor{\bsnm{{Shih}}, \binits{A.Y.}},
\bauthor{\bsnm{{Lin}}, \binits{R.P.}},
\bauthor{\bsnm{{Smith}}, \binits{D.M.}}:
\byear{2009},
\batitle{{RHESSI Observations of the Proportional Acceleration of Relativistic >0.3 MeV Electrons and >30 MeV Protons in Solar Flares}}.
\bjtitle{\apjl}
\bvolume{698},
\bfpage{L152}.
\doiurl{https://doi.org/10.1088/0004-637X/698/2/L152}.
\adsurl{2009ApJ...698L.152S}.
\end{barticle}
\endbibitem

\bibitem[\protect\citeauthoryear{{Veronig} et~al.}{2010}]{veronig2010}
\begin{barticle}
\bauthor{\bsnm{{Veronig}}, \binits{A.M.}},
\bauthor{\bsnm{{Muhr}}, \binits{N.}},
\bauthor{\bsnm{{Kienreich}}, \binits{I.W.}},
\bauthor{\bsnm{{Temmer}}, \binits{M.}},
\bauthor{\bsnm{{Vr{\v{s}}nak}}, \binits{B.}}:
\byear{2010},
\batitle{{First Observations of a Dome-shaped Large-scale Coronal Extreme-ultraviolet Wave}}.
\bjtitle{\apjl}
\bvolume{716},
\bfpage{L57}.
\doiurl{https://doi.org/10.1088/2041-8205/716/1/L57}.
\adsurl{2010ApJ...716L..57V}.
\end{barticle}
\endbibitem

\bibitem[\protect\citeauthoryear{{Vestrand} and {Forrest}}{1993}]{Vestrand1993}
\begin{barticle}
\bauthor{\bsnm{{Vestrand}}, \binits{W.T.}},
\bauthor{\bsnm{{Forrest}}, \binits{D.J.}}:
\byear{1993},
\batitle{{Evidence for a Spatially Extended Component of Gamma Rays from Solar Flares}}.
\bjtitle{\apjl}
\bvolume{409},
\bfpage{L69}.
\doiurl{https://doi.org/10.1086/186862}.
\adsurl{1993ApJ...409L..69V}.
\end{barticle}
\endbibitem

\bibitem[\protect\citeauthoryear{{Vilmer}, {MacKinnon}, and {Hurford}}{2011}]{Vilmer2011}
\begin{barticle}
\bauthor{\bsnm{{Vilmer}}, \binits{N.}},
\bauthor{\bsnm{{MacKinnon}}, \binits{A.L.}},
\bauthor{\bsnm{{Hurford}}, \binits{G.J.}}:
\byear{2011},
\batitle{{Properties of Energetic Ions in the Solar Atmosphere from {\ensuremath{\gamma}}-Ray and Neutron Observations}}.
\bjtitle{\ssr}
\bvolume{159},
\bfpage{167}.
\doiurl{https://doi.org/10.1007/s11214-010-9728-x}.
\adsurl{2011SSRv..159..167V}.
\end{barticle}
\endbibitem

\bibitem[\protect\citeauthoryear{{Vr{\v{s}}nak} and {Luli{\'c}}}{2000}]{Vrsnak2000}
\begin{barticle}
\bauthor{\bsnm{{Vr{\v{s}}nak}}, \binits{B.}},
\bauthor{\bsnm{{Luli{\'c}}}, \binits{S.}}:
\byear{2000},
\batitle{{Formation Of Coronal Mhd Shock Waves - I. The Basic Mechanism}}.
\bjtitle{\solphys}
\bvolume{196},
\bfpage{157}.
\doiurl{https://doi.org/10.1023/A:1005236804727}.
\adsurl{2000SoPh..196..157V}.
\end{barticle}
\endbibitem

\bibitem[\protect\citeauthoryear{{Vr{\v{s}}nak} et~al.}{2001}]{vrsnak2001}
\begin{barticle}
\bauthor{\bsnm{{Vr{\v{s}}nak}}, \binits{B.}},
\bauthor{\bsnm{{Aurass}}, \binits{H.}},
\bauthor{\bsnm{{Magdaleni{\'c}}}, \binits{J.}},
\bauthor{\bsnm{{Gopalswamy}}, \binits{N.}}:
\byear{2001},
\batitle{{Band-splitting of coronal and interplanetary type II bursts. I. Basic properties}}.
\bjtitle{\aap}
\bvolume{377},
\bfpage{321}.
\doiurl{https://doi.org/10.1051/0004-6361:20011067}.
\adsurl{2001A&A...377..321V}.
\end{barticle}
\endbibitem

\bibitem[\protect\citeauthoryear{{Warmuth}}{2010}]{Warmuth2010}
\begin{barticle}
\bauthor{\bsnm{{Warmuth}}, \binits{A.}}:
\byear{2010},
\batitle{{Large-scale waves in the solar corona: The continuing debate}}.
\bjtitle{Advances in Space Research}
\bvolume{45},
\bfpage{527}.
\doiurl{https://doi.org/10.1016/j.asr.2009.08.022}.
\adsurl{2010AdSpR..45..527W}.
\end{barticle}
\endbibitem

\bibitem[\protect\citeauthoryear{{Warmuth} and {Mann}}{2005}]{Warmuth2005}
\begin{barticle}
\bauthor{\bsnm{{Warmuth}}, \binits{A.}},
\bauthor{\bsnm{{Mann}}, \binits{G.}}:
\byear{2005},
\batitle{{A model of the Alfv{\'e}n speed in the solar corona}}.
\bjtitle{\aap}
\bvolume{435},
\bfpage{1123}.
\doiurl{https://doi.org/10.1051/0004-6361:20042169}.
\adsurl{2005A&A...435.1123W}.
\end{barticle}
\endbibitem

\bibitem[\protect\citeauthoryear{{Warmuth} and {Mann}}{2011}]{2011A&A...532A.151W}
\begin{barticle}
\bauthor{\bsnm{{Warmuth}}, \binits{A.}},
\bauthor{\bsnm{{Mann}}, \binits{G.}}:
\byear{2011},
\batitle{{Kinematical evidence for physically different classes of large-scale coronal EUV waves}}.
\bjtitle{\aap}
\bvolume{532},
\bfpage{A151}.
\doiurl{https://doi.org/10.1051/0004-6361/201116685}.
\adsurl{2011A&A...532A.151W}.
\end{barticle}
\endbibitem

\bibitem[\protect\citeauthoryear{{Warmuth} and {Mann}}{2020}]{Warmuth2020}
\begin{barticle}
\bauthor{\bsnm{{Warmuth}}, \binits{A.}},
\bauthor{\bsnm{{Mann}}, \binits{G.}}:
\byear{2020},
\batitle{{Thermal-nonthermal energy partition in solar flares derived from X-ray, EUV, and bolometric observations. Discussion of recent studies}}.
\bjtitle{\aap}
\bvolume{644},
\bfpage{A172}.
\doiurl{https://doi.org/10.1051/0004-6361/202039529}.
\adsurl{2020A&A...644A.172W}.
\end{barticle}
\endbibitem

\bibitem[\protect\citeauthoryear{{Warmuth} et~al.}{2004}]{warmuth2004b}
\begin{barticle}
\bauthor{\bsnm{{Warmuth}}, \binits{A.}},
\bauthor{\bsnm{{Vr{\v{s}}nak}}, \binits{B.}},
\bauthor{\bsnm{{Magdaleni{\'c}}}, \binits{J.}},
\bauthor{\bsnm{{Hanslmeier}}, \binits{A.}},
\bauthor{\bsnm{{Otruba}}, \binits{W.}}:
\byear{2004},
\batitle{{A multiwavelength study of solar flare waves. II. Perturbation characteristics and physical interpretation}}.
\bjtitle{\aap}
\bvolume{418},
\bfpage{1117}.
\doiurl{https://doi.org/10.1051/0004-6361:20034333}.
\adsurl{2004A&A...418.1117W}.
\end{barticle}
\endbibitem

\bibitem[\protect\citeauthoryear{{White} et~al.}{2011}]{White2011}
\begin{barticle}
\bauthor{\bsnm{{White}}, \binits{S.M.}},
\bauthor{\bsnm{{Benz}}, \binits{A.O.}},
\bauthor{\bsnm{{Christe}}, \binits{S.}},
\bauthor{\bsnm{{F{\'a}rn{\'\i}k}}, \binits{F.}},
\bauthor{\bsnm{{Kundu}}, \binits{M.R.}},
\bauthor{\bsnm{{Mann}}, \binits{G.}},
\bauthor{\bsnm{{Ning}}, \binits{Z.}},
\bauthor{\bsnm{{Raulin}}, \binits{J.-P.}},
\bauthor{\bsnm{{Silva-V{\'a}lio}}, \binits{A.V.R.}},
\bauthor{\bsnm{{Saint-Hilaire}}, \binits{P.}},
\bauthor{\bsnm{{Vilmer}}, \binits{N.}},
\bauthor{\bsnm{{Warmuth}}, \binits{A.}}:
\byear{2011},
\batitle{{The Relationship Between Solar Radio and Hard X-ray Emission}}.
\bjtitle{\ssr}
\bvolume{159},
\bfpage{225}.
\doiurl{https://doi.org/10.1007/s11214-010-9708-1}.
\adsurl{2011SSRv..159..225W}.
\end{barticle}
\endbibitem

\bibitem[\protect\citeauthoryear{{Wild} and {McCready}}{1950}]{wild1950}
\begin{barticle}
\bauthor{\bsnm{{Wild}}, \binits{J.P.}},
\bauthor{\bsnm{{McCready}}, \binits{L.L.}}:
\byear{1950},
\batitle{Observations of the Spectrum of High-Intensity Solar Radiation at Metre Wavelengths. I. The Apparatus and Spectral Types of Solar Burst Observed}.
\bjtitle{Aust. J. Sci. Res. A}
\bvolume{3},
\bfpage{387}.
\adsurl{http://esoads.eso.org/abs/1950AuSRA...3..387W}.
\end{barticle}
\endbibitem

\bibitem[\protect\citeauthoryear{{Wuelser} et~al.}{2004}]{wuelser2004}
\begin{bchapter}
\bauthor{\bsnm{{Wuelser}}, \binits{J.-P.}},
\bauthor{\bsnm{{Lemen}}, \binits{J.R.}},
\bauthor{\bsnm{{Tarbell}}, \binits{T.D.}},
\bauthor{\bsnm{{Wolfson}}, \binits{C.J.}},
\bauthor{\bsnm{{Cannon}}, \binits{J.C.}},
\bauthor{\bsnm{{Carpenter}}, \binits{B.A.}},
\bauthor{\bsnm{{Duncan}}, \binits{D.W.}},
\bauthor{\bsnm{{Gradwohl}}, \binits{G.S.}},
\bauthor{\bsnm{{Meyer}}, \binits{S.B.}},
\bauthor{\bsnm{{Moore}}, \binits{A.S.}},
\bauthor{\bsnm{{Navarro}}, \binits{R.L.}},
\bauthor{\bsnm{{Pearson}}, \binits{J.D.}},
\bauthor{\bsnm{{Rossi}}, \binits{G.R.}},
\bauthor{\bsnm{{Springer}}, \binits{L.A.}},
\bauthor{\bsnm{{Howard}}, \binits{R.A.}},
\bauthor{\bsnm{{Moses}}, \binits{J.D.}},
\bauthor{\bsnm{{Newmark}}, \binits{J.S.}},
\bauthor{\bsnm{{Delaboudiniere}}, \binits{J.-P.}},
\bauthor{\bsnm{{Artzner}}, \binits{G.E.}},
\bauthor{\bsnm{{Auchere}}, \binits{F.}},
\bauthor{\bsnm{{Bougnet}}, \binits{M.}},
\bauthor{\bsnm{{Bouyries}}, \binits{P.}},
\bauthor{\bsnm{{Bridou}}, \binits{F.}},
\bauthor{\bsnm{{Clotaire}}, \binits{J.-Y.}},
\bauthor{\bsnm{{Colas}}, \binits{G.}},
\bauthor{\bsnm{{Delmotte}}, \binits{F.}},
\bauthor{\bsnm{{Jerome}}, \binits{A.}},
\bauthor{\bsnm{{Lamare}}, \binits{M.}},
\bauthor{\bsnm{{Mercier}}, \binits{R.}},
\bauthor{\bsnm{{Mullot}}, \binits{M.}},
\bauthor{\bsnm{{Ravet}}, \binits{M.-F.}},
\bauthor{\bsnm{{Song}}, \binits{X.}},
\bauthor{\bsnm{{Bothmer}}, \binits{V.}},
\bauthor{\bsnm{{Deutsch}}, \binits{W.}}:
\byear{2004},
\bctitle{{EUVI: the STEREO-SECCHI extreme ultraviolet imager}}.
In: \beditor{\bsnm{{Fineschi}}, \binits{S.}},
\beditor{\bsnm{{Gummin}}, \binits{M.A.}} (eds.)
\bbtitle{Telescopes and Instrumentation for Solar Astrophysics},
\bsertitle{Society of Photo-Optical Instrumentation Engineers (SPIE) Conference Series}
\bseriesno{5171},
\bfpage{111}.
\doiurl{https://doi.org/10.1117/12.506877}.
\adsurl{2004SPIE.5171..111W}.
\end{bchapter}
\endbibitem

\bibitem[\protect\citeauthoryear{{Yashiro} et~al.}{2004}]{Yashiro2004}
\begin{barticle}
\bauthor{\bsnm{{Yashiro}}, \binits{S.}},
\bauthor{\bsnm{{Gopalswamy}}, \binits{N.}},
\bauthor{\bsnm{{Michalek}}, \binits{G.}},
\bauthor{\bsnm{{St. Cyr}}, \binits{O.C.}},
\bauthor{\bsnm{{Plunkett}}, \binits{S.P.}},
\bauthor{\bsnm{{Rich}}, \binits{N.B.}},
\bauthor{\bsnm{{Howard}}, \binits{R.A.}}:
\byear{2004},
\batitle{{A catalog of white light coronal mass ejections observed by the SOHO spacecraft}}.
\bjtitle{Journal of Geophysical Research (Space Physics)}
\bvolume{109},
\bfpage{A07105}.
\doiurl{https://doi.org/10.1029/2003JA010282}.
\adsurl{2004JGRA..109.7105Y}.
\end{barticle}
\endbibitem

\bibitem[\protect\citeauthoryear{{Zucca} et~al.}{2018}]{Zucca2018}
\begin{barticle}
\bauthor{\bsnm{{Zucca}}, \binits{P.}},
\bauthor{\bsnm{{Morosan}}, \binits{D.E.}},
\bauthor{\bsnm{{Rouillard}}, \binits{A.P.}},
\bauthor{\bsnm{{Fallows}}, \binits{R.}},
\bauthor{\bsnm{{Gallagher}}, \binits{P.T.}},
\bauthor{\bsnm{{Magdalenic}}, \binits{J.}},
\bauthor{\bsnm{{Klein}}, \binits{K.-L.}},
\bauthor{\bsnm{{Mann}}, \binits{G.}},
\bauthor{\bsnm{{Vocks}}, \binits{C.}},
\bauthor{\bsnm{{Carley}}, \binits{E.P.}},
\bauthor{\bsnm{{Bisi}}, \binits{M.M.}},
\bauthor{\bsnm{{Kontar}}, \binits{E.P.}},
\bauthor{\bsnm{{Rothkaehl}}, \binits{H.}},
\bauthor{\bsnm{{Dabrowski}}, \binits{B.}},
\bauthor{\bsnm{{Krankowski}}, \binits{A.}},
\bauthor{\bsnm{{Anderson}}, \binits{J.}},
\bauthor{\bsnm{{Asgekar}}, \binits{A.}},
\bauthor{\bsnm{{Bell}}, \binits{M.E.}},
\bauthor{\bsnm{{Bentum}}, \binits{M.J.}},
\bauthor{\bsnm{{Best}}, \binits{P.}},
\bauthor{\bsnm{{Blaauw}}, \binits{R.}},
\bauthor{\bsnm{{Breitling}}, \binits{F.}},
\bauthor{\bsnm{{Broderick}}, \binits{J.W.}},
\bauthor{\bsnm{{Brouw}}, \binits{W.N.}},
\bauthor{\bsnm{{Br{\"u}ggen}}, \binits{M.}},
\bauthor{\bsnm{{Butcher}}, \binits{H.R.}},
\bauthor{\bsnm{{Ciardi}}, \binits{B.}},
\bauthor{\bsnm{{de Geus}}, \binits{E.}},
\bauthor{\bsnm{{Deller}}, \binits{A.}},
\bauthor{\bsnm{{Duscha}}, \binits{S.}},
\bauthor{\bsnm{{Eisl{\"o}ffel}}, \binits{J.}},
\bauthor{\bsnm{{Garrett}}, \binits{M.A.}},
\bauthor{\bsnm{{Grie{\ss}meier}}, \binits{J.M.}},
\bauthor{\bsnm{{Gunst}}, \binits{A.W.}},
\bauthor{\bsnm{{Heald}}, \binits{G.}},
\bauthor{\bsnm{{Hoeft}}, \binits{M.}},
\bauthor{\bsnm{{H{\"o}randel}}, \binits{J.}},
\bauthor{\bsnm{{Iacobelli}}, \binits{M.}},
\bauthor{\bsnm{{Juette}}, \binits{E.}},
\bauthor{\bsnm{{Karastergiou}}, \binits{A.}},
\bauthor{\bsnm{{van Leeuwen}}, \binits{J.}},
\bauthor{\bsnm{{McKay-Bukowski}}, \binits{D.}},
\bauthor{\bsnm{{Mulder}}, \binits{H.}},
\bauthor{\bsnm{{Munk}}, \binits{H.}},
\bauthor{\bsnm{{Nelles}}, \binits{A.}},
\bauthor{\bsnm{{Orru}}, \binits{E.}},
\bauthor{\bsnm{{Paas}}, \binits{H.}},
\bauthor{\bsnm{{Pandey}}, \binits{V.N.}},
\bauthor{\bsnm{{Pekal}}, \binits{R.}},
\bauthor{\bsnm{{Pizzo}}, \binits{R.}},
\bauthor{\bsnm{{Polatidis}}, \binits{A.G.}},
\bauthor{\bsnm{{Reich}}, \binits{W.}},
\bauthor{\bsnm{{Rowlinson}}, \binits{A.}},
\bauthor{\bsnm{{Schwarz}}, \binits{D.J.}},
\bauthor{\bsnm{{Shulevski}}, \binits{A.}},
\bauthor{\bsnm{{Sluman}}, \binits{J.}},
\bauthor{\bsnm{{Smirnov}}, \binits{O.}},
\bauthor{\bsnm{{Sobey}}, \binits{C.}},
\bauthor{\bsnm{{Soida}}, \binits{M.}},
\bauthor{\bsnm{{Thoudam}}, \binits{S.}},
\bauthor{\bsnm{{Toribio}}, \binits{M.C.}},
\bauthor{\bsnm{{Vermeulen}}, \binits{R.}},
\bauthor{\bsnm{{van Weeren}}, \binits{R.J.}},
\bauthor{\bsnm{{Wucknitz}}, \binits{O.}},
\bauthor{\bsnm{{Zarka}}, \binits{P.}}:
\byear{2018},
\batitle{{Shock location and CME 3D reconstruction of a solar type II radio burst with LOFAR}}.
\bjtitle{\aap}
\bvolume{615},
\bfpage{A89}.
\doiurl{https://doi.org/10.1051/0004-6361/201732308}.
\adsurl{2018A&A...615A..89Z}.
\end{barticle}
\endbibitem

\end{thebibliography}

\IfFileExists{\jobname.bbl}{} {\typeout{}
\typeout{****************************************************}
\typeout{****************************************************}
\typeout{** Please run "bibtex \jobname" to obtain} \typeout{**
the bibliography and then re-run LaTeX} \typeout{** twice to fix
the references !}
\typeout{****************************************************}
\typeout{****************************************************}
\typeout{}}

\end{document}